\documentclass[preprint,12pt]{elsarticle}

\usepackage{amssymb}
\usepackage{amsmath}

\usepackage{url}
\usepackage{graphicx}
\usepackage{multirow}
\usepackage{multicol}
\usepackage{lscape}
\usepackage{hyperref}
\usepackage{balance}
\usepackage{todonotes}
\usepackage{xcolor}
\usepackage{array}

\journal{Nuclear Physics B}

\begin{document}

\begin{frontmatter}

\title{Taming the Online Software Engineering Tongue: Detecting and
Reducing Offensive Language in Software Communities} 

\author{Jithin Cheriyan}
\author{Bastin Tony Roy Savarimuthu}
\author{Stephen Cranefield}

\affiliation{organization={School of Computing, University of Otago}, 
            city={Dunedin},
            country={New Zealand}}

\begin{abstract}
Online Software Engineering (SE) communities such as Stack Overflow have become unwelcoming and hostile spaces, as evidenced by the presence of rude and offensive comments that are detrimental to the emotional well-being of participants. The main objective of this work is to investigate the prevalence and nature of offensive language in online SE communities, and develop approaches to classify them with the ultimate goal of reducing offensive language by proposing an Offensive Language Reduction System (OLRS-SE) that has the ability to detect, classify, and explain to the poster why a comment is offensive and also propose non-offensive alternative suggestions. Using data from 130,000 user comments on discussions held in four prominent SE platforms: Stack Overflow, GitHub, Gitter and Slack, we developed six sets of twelve (72 in total) different machine learning models (both traditional and deep learning). The F1-Score of the best performing models in the system ranges from 0.56 to 0.94, with an average F1-Score of 0.85, predominantly using BERT as the primary model for various classification tasks. Also, two language-based models were developed for paraphrasing offensive to non-offensive comments. The overall utility of the OLRS-SE was evaluated using user studies with 30 participants, which showed that the developed system is easy to use and also effective in reducing offensive language. Investigating the capabilities of different large language models for offensive language reduction is a promising avenue for future research.
\end{abstract}


\end{frontmatter}

\section{Introduction}

  Online social networking platforms such as Facebook and Twitter have created opportunities for individuals to interact within virtual communities. In the software development domain, Stack Overflow (SO)\footnote{Stack Overflow: \url{https://stackoverflow.com}}, is a platform  well-known for its technical question-answering and problem-solving support for professionals and enthusiast  programmers. Similar online SE platforms include GitHub\footnote{GitHub: \url{https://github.com}}, Gitter\footnote{Gitter: \url{https://gitter.im}} and Slack\footnote{Slack: \url{https://slack.com}}. Interactions between users of these applications are generally observed as professional, positive and inclusive~\citep{jones_online_2013}. The Code of Conduct (CoC)\footnote{Stack Overflow CoC: \url{https://stackoverflow.com/conduct}} of SO calls for its members to \textit{be inclusive and respectful}. However, hate speech and abusive language usage have increasingly been observed on the rise~\citep{jones_online_2013, DBLP:conf/gldv/BourgonjeSSR17, ALW_2019}. 

    SO archive\footnote{Stack Overflow archive: \url{https://meta.stackoverflow.com/questions/326494}} maintains sample offensive comments, which reveals the presence of offence and hostility in the platform\footnote{\textit{Disclaimer}: This work contains comments that carry strong language or profanity, which may be offensive or upsetting to some readers. Hence, reader discretion is advised.}. This normative shift of SO to unwelcoming and less inclusive creates mental health issues~\citep{Raman_2020} that can cause members to leave the platform~\citep{so_culture_stack_2018}. Moreover, newcomers face hurdles on the site such as negligence of posts, deleting or closing of posts, and downvoting from others without proper reasoning~\citep{DBLP:conf/enase/JobairMMMH22}. This gradual onset of impoliteness has been admitted by the former executive Vice President of SO who noted that ``\textit{too many people experience Stack Overflow as a hostile or elitist place, especially newer coders, women, people of color, and others in marginalized groups}''~\citep{so_culture_stack_2018}. Thus, investigating the prevalence of offensive language, their classes, and exploring approaches that can reduce such offences on online SE platforms form the context of this work. To address this problem, we propose a novel framework termed Offensive Language Reduction System for Software Engineering (OLRS-SE) which can \textit{detect} and \textit{classify} offence,  \textit{explain} the reason for offensiveness in a comment and \textit{propose} potential non-offensive suggestions for an offensive comment. Therefore, the specific research questions that we address in this work are:

    \noindent \textbf{RQ1: What is the nature and prevalence of offensive language in SE communities?}\\
    \textbf{RQ2: How can an Offensive Language Reduction System for Software Engineering (OLRS-SE) be developed? } \\
    \hspace*{10mm}\textbf{RQ2.1: How can we develop approaches to accurately \textit{detect} and \textit{classify} offensive language in online SE communities, using machine learning techniques?} \\
    \hspace*{10mm}\textbf{RQ2.2: How can we develop approaches to \textit{explain} offensive language to its poster and also \textit{rephrase} offensive ones into less offensive ones using language models?}\\

    The paper is structured as follows. Section~\ref{sec:methodology} presents the methodology for detecting and reducing offensive language in SE communities by proposing the recommendation system. The results for RQ1 and RQ2 are presented in Section~\ref{sec:results}. The utility of the developed system is evaluated in Section~\ref{sec:user_evluation}. Section~\ref{sec:discussion} discusses the implications of the results, Section~\ref{sec:related_work} discusses related work on the detection and classification of offensive language in SE communities, and Section~\ref{sec:conclusion} concludes the paper.

\section{Methodology}
\label{sec:methodology}
    This section describes the approaches that we used to detect the nature and prevalence of offensive language (RQ1), and develop the OLRS-SE framework to detect, classify, explain and reduce offensive language in SE communities (RQ2). 
    
    \subsection{Nature and prevalence of offensive language (RQ1)}

    \subsubsection{Data collection}
    \label{subsection:data_collection}
    To identify the nature and prevalence of offensive language in SE communities (RQ1), we obtained the dataset of deleted comments from SO obtained in 2020, which was released as part of the Workshop on Online Abuse and Harms (WOAH-2020)\footnote{WOAH: \url{https://sites.google.com/view/abusiveworkshop}}.

        Of the 311073 comments with five classes, the SO dataset contained two offence classes which we consider in this work: Rude/Offensive (\textit{RO}) and Unfriendly/Unwelcoming (\textit{UN}).  RO comments (108473, 34.87\%)  were flagged by users or the SO automatic flagging system (a bot) and have also been judged rude or offensive by human moderators. Those comments were subsequently removed from the site as part of the moderation, as these comments are highly offensive or toxic in nature. The UN comments (28030, 9.01\%) are comments with less offensive content than RO, but through the flagging and moderation process, they were also removed from the site because they were deemed unfriendly. Hence, these two classes of comments contain offensive language, as they elicited sanctions for breaching the CoC of the site (i.e., they were deleted from the site). We treat both RO and UN labels as identifying unwelcoming comments (i.e.,  offensive and unfriendly comments, respectively). 

    \subsubsection{Taxonomy generation}
    \label{subsection:taxonomy}
        This section describes a taxonomy we have developed which is  used to classify offensive language into nuanced classes, thus revealing the nature  of these offensive comments. Quantification of these classes shows the prevalence of offensive language in SE communities, thus answering RQ1. The objective of the nuanced classification is to better understand and identify the target of the offensive language users (i.e., those who violate the CoC), and communicate these violations to the writers. To achieve that, we extended the taxonomy presented by Cheriyan et al.~\citep{DBLP:conf/coin-ws/CheriyanSC20} to identify different fine-grained classes of offensive language. Using the RO and UN comments dataset, and employing a combination of bottom-up and top-down approaches~\citep{DBLP:conf/icse/CapiluppiA20}, we constructed the taxonomy. The top-down approach applies existing taxonomy attributes to the dataset while the bottom-up approach unearths new attributes from the dataset~\citep{DBLP:conf/icse/CapiluppiA20}. Using these approaches, we identified the attributes from a random selection of 500 comments each from RO and UN datasets. The first author identified the attributes and the second author validated them (i.e., full consensus was achieved).  Moreover, we followed a layered approach where in each layer, fine-grained attributes of an offensive comment are further incrementally unpacked similar to the approach in~\citep{DBLP:conf/icis/PuschelRS16}. Table \ref{table:Fine_grained_taxonomy} presents the detailed taxonomy that classifies the offensive language into three different levels, namely, Level-1, Level-2 and Level-3. It also presents examples of offensive comments in the last column. Note that the taxonomy presented in~\citep{DBLP:conf/coin-ws/CheriyanSC20} has just one level, and ours has three. Also~\citep{DBLP:conf/coin-ws/CheriyanSC20} considers only eight attributes, compared to 14 attributes in our work. Also, the work in~\citep{DBLP:conf/coin-ws/CheriyanSC20} uses a keyword based approach to identify attributes while we employ machine learning based approaches in this work. A hierarchical representation of our taxonomy can be found online\footnote{Detailed taxonomy: \url{https://doi.org/10.6084/m9.figshare.23691618}}.

         \begin{table}        
           \begin{tabular}{|p{1.5cm}|p{3cm}|p{3cm}|p{4cm}|}
                \hline
                \textbf{Level-1} & \textbf{Level-2} & \textbf{Level-3} & \textbf{Example}\\ 
                \hline
            
            \multirow{17}{1.25cm} {Explicit offence (RO)~\citep{DBLP:conf/icwsm/DinakarRL11,waseem2017understanding,wiegand-etal-2018-inducing, zampieri-etal-2019-predicting} \\ (29381)} 
                    & \multirow{6}{3cm}{Individually targeted~\citep{waseem2017understanding, zampieri-etal-2019-predicting}\\ (15692)}
                        & \multirow{3}{3cm} {Racial~\citep{waseem2017understanding,DBLP:conf/gldv/BourgonjeSSR17}\\ (1040)} 
                            & \textit{Try VS code nigga!!}\\\cline{4-4}
                         & & & \textit{Retarded Indian IT idiot?}\\\cline{3-4}
                        
                        &  & \multirow{3}{3cm}{Swearing~\citep{waseem2017understanding,DBLP:conf/gldv/BourgonjeSSR17}\\ (14658)}
                            & \textit{F** you @Andy.}\\\cline{4-4}
                        & & & \textit{@User, I call b**** on your answer.}\\\cline{2-4}
                        
                    & \multirow{7}{3cm}{Generalized~\citep{DBLP:conf/icwsm/DinakarRL11, waseem2017understanding, zampieri-etal-2019-predicting}\\ (5716)} 
                        & \multirow{4}{3cm}{Racial~\citep{waseem2017understanding,DBLP:conf/gldv/BourgonjeSSR17}\\ (333)}
                            & \textit{Niggas using Wamp Server…?}\\\cline{4-4}
                        & & & \textit{Block all F***ing Chinese.} \\\cline{3-4}
                        
                        & & \multirow{3}{3cm}{Swearing~\citep{waseem2017understanding,DBLP:conf/gldv/BourgonjeSSR17}\\ (3187)} 
                             & \textit{WTF programmers?}\\\cline{4-4}
                        &  &  & \textit{Who the f** downvoted?}\\\cline{2-4}

                    & Entity abuse (644) & \multirow{2}{3cm} {Swearing \\(644)} & \textit{Python sucks!!!}\\\cline{2-4}
                        
                    & \multirow{2}{3cm} {Self-abuse \\(263)} & \multirow{2}{3cm} {Swearing \\(263)} & \textit{You are right, stupid me.}\\\cline{2-4}

                    & Others~\citep{mondal2017measurement, zampieri-etal-2019-predicting} (7410) & \multirow{2}{3cm} {Swearing\\  (7410)} & \textit{Seriously.. F***k off}\\\cline{1-4}

            \multirow{14}{1.5cm} {Milder offence (UN)~\citep{DBLP:conf/icwsm/DinakarRL11,waseem2017understanding,wiegand-etal-2018-inducing} \\ (6879)} 
                    & \multirow{7}{3cm}{Individually targeted~\citep{waseem2017understanding} \\(6100)} 
                         & \multirow{3}{3cm}{Racial~\citep{waseem2017understanding,DBLP:conf/gldv/BourgonjeSSR17} }
                            & \textit{That's a nice example of a bad Indian question.}\\\cline{3-4}
                        
                        &  & \multirow{3}{3cm}{Swearing~\citep{waseem2017understanding,DBLP:conf/gldv/BourgonjeSSR17} }
                            & \textit{Are you making your dog solve the problems for you ???} \\\cline{2-4}
                      
                    & \multirow{5}{3cm}{Generalized~\citep{DBLP:conf/icwsm/DinakarRL11, waseem2017understanding}\\ (457)} 
                        & \multirow{3}{3cm}{Racial~\citep{waseem2017understanding,DBLP:conf/gldv/BourgonjeSSR17} }
                         &  \textit{OMG... like the Mexican police problem solving..?}\\\cline{3-4}
                        
                        &  & \multirow{2}{3cm}{Swearing~\citep{waseem2017understanding,DBLP:conf/gldv/BourgonjeSSR17} }
                         &  \textit{You people need a rubber duck my friends...}\\\cline{2-4}

                    & Entity abuse (60) & \multirow{5}{3cm}{NA} & \textit{Java is a nanny language.}\\\cline{2-2} \cline{4-4}
                        
                    & Self-abuse (2)  &  & \textit{I am LAZY, I can't!!!}\\\cline{2-2}\cline{4-4}

                    & \multirow{2}{3cm} {Others~\citep{mondal2017measurement}\\  (358)} &  & \textit{SO isn't a code writing service.}\\\cline{1-4}
            
            \end{tabular}
            
           \caption{Taxonomy of fine-grained offence classification (the numbers in each class in red represents the number of comments that belongs to that class of offence)}
           \label{table:Fine_grained_taxonomy}
        \end{table}
        
        In Level-1, the presence of offence in a comment is identified. As per SO community labelling mentioned in Section~\ref{subsection:data_collection}, an offensive comment may be either RO or UN. The RO comments are explicitly offensive comments bearing highly toxic contents. Since the RO comments are explicit offences, henceforth, we use these two terms (i.e., RO and explicit offence) interchangeably. However, the UN comments hold less offensive contents than RO comments. Even though the toxicity boundary between RO and UN is narrow, there exists a border line between these two classes. For example, ``\textit{I hate f***ing java code}" is an RO comment, while ``\textit{SO isn’t a code writing service. You should do your homework yourself}" is a UN comment. In order to decide whether a comment is offensive, we need to identify whether that comment is either RO or UN. A positive outcome for either of these classes identifies a comment as offensive. Therefore, the detection of these two classes (RO and UN) together constitutes the offence identification in Level-1. If a comment is not identified as either RO or UN, that would be considered non-offensive.

        In Level-2, if a comment has been identified as offensive (either RO or UN), the target of the offence is identified (e.g., an individual or a group). By identifying and detecting the target of the comment (e.g., an individual or a group), the target of the comment posters can be identified. Based on prior literature and data coding (i.e., use of top-down and bottom-up approaches~\citep{DBLP:conf/icse/CapiluppiA20}), we identified the classes in Level-2 of the taxonomy, which are \textit{Individually targeted}, \textit{Generalized}, \textit{Self-abuse}, \textit{Entity abuse} and \textit{Others}. Examples for these classes are provided in the fourth column of Table~\ref{table:Fine_grained_taxonomy}. An individually targeted offence is a comment intended to abuse a particular individual by mentioning some personal identifier (e.g., user name)~\citep{waseem2017understanding}. A generalized or group-targeted offence is a comment targeting people belonging to a particular category or group (e.g., country, race, religion, sexual orientation or administrators of a site) (\cite{mondal2017measurement, waseem2017understanding}). \textit{Entity abuse} targets some technology (e.g., cloud service), software or hardware product (e.g., Windows, VB.Net) or platform (e.g., SO, GitHub). This class of offence consists of comments that abuse or express negative opinion about a technology, a product or a system that is worded in an offensive way. \textit{Self-abuse} is the class of offence where authors of the comments target themselves by mocking or cursing. For example, comments such as ``\textit{Oh!!! I F***d myself by that code}'' are cursing or targeting themselves. Even though these may be perceived to be innocuous in nature, these comments can still be offensive to readers~\citep{mondal2017measurement}. If the target of offence does not belong to any of the above-mentioned classes and the intention of the writer is unclear, it is classified as \textit{Others}~\citep{mondal2017measurement}.
        
        In Level-3, classes representing two specific reasons why the content is deemed offensive are identified. Two common reasons are the use of \textit{Racial} and \textit{Swearing} terms (\cite{waseem2017understanding,DBLP:conf/gldv/BourgonjeSSR17}). Use of terms in a comment that abuses one's ethnicity, skin tone or any other kind of racial discrimination is labelled as \textit{Racial}. Use of swear words is labelled as \textit{Swearing}. 
        
        It should be noted that a comment can contain multiple labels within levels 2 and 3. For example, a comment can contain both swearing terms and racially offensive terms.

        The quantification of comments in each class is also mentioned in Table \ref{table:Fine_grained_taxonomy} (see the numbers in red). These are the number of comments after removing non-English comments, comments with emojis alone and duplicates, and these numbers were obtained based on manually reading the comments by the first author. The second author validated the results by checking a random sample of 200 comments. This result shows the prevalence of different classes of offensive comments in SE communities. There were substantially more RO comments than UN comments in Level-1. Also, there were more individually targeted comments the other types in Level-2. Additionally, there were more swearing comments than racial comments in Level-3. Owing to multi-labelling, the sum of comments in Level-2 and Level-3 may exceed that of Level-1. Since UN comments are mild offences, they do not contain racial or swearing words. So, we did classify UN comments at Level-3 - i.e., NA (not applicable) is assigned. The rubric used for offence labelling can be found online\footnote{Rubric for offence labelling: \url{https://doi.org/10.6084/m9.figshare.27129693}}.

    \subsection{OLRS-SE framework (RQ2)}
    \label{subsection:OLRS-SE_framework}

\begin{figure}[h]
        \includegraphics[width=\linewidth]{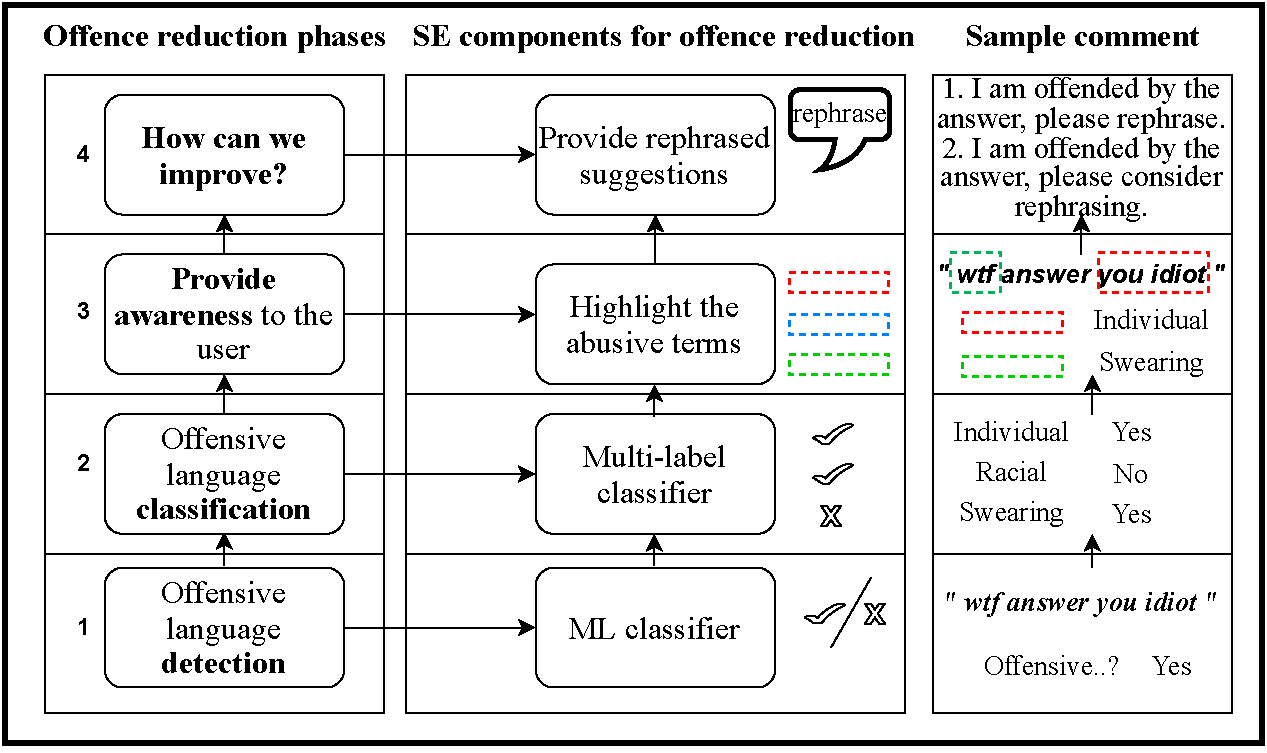}
        \caption{OLRS-SE framework}
        \label{fig:OLRS-SE_framework}
        \end{figure}

        To address RQ2, we propose OLRS-SE, a hybrid system that incorporates rule-based and machine learning (ML) approaches. As shown in Figure~\ref{fig:OLRS-SE_framework}, this is a framework for offensive language detection and reduction in SE platforms. An offensive comment entered by a user will be processed using a four-phased approach. The first phase employs a binary classifier to detect whether a comment is offensive or not (corresponding to Level-1 of the taxonomy in Table \ref{table:Fine_grained_taxonomy}), by using ML approaches (discussed in Section~\ref{subsection:ML_model_building}).  
        
        The second phase employs a multi-label classifier to classify the offence in a fine-grained fashion (corresponding to Levels 2 and 3 of the taxonomy in Table \ref{table:Fine_grained_taxonomy}). The third phase highlights the offensive contents to the writer and mentions the type of offence to the writer (e.g., racial). Once a comment has been classified as offensive, a machine learning based  explanation system highlights the offensive  terms (i.e., using dashed lines around the offensive terms), which are the reasons for a comment being labelled as an offence. This has been identified using ML explanation methods and is addressed in Section~\ref{subsection_offence_rephrasing}.
        
        Even after being informed about the offence, if the writers like to express disagreement with a post, they are offered opportunities to express this without using offensive language. This final phase involves text detoxification using text paraphrasing approaches. In this phase, the system provides a set of paraphrased statements which eliminate or substitute the offensive terms with non-offensive ones. As a part of this process, we do not change the tone of the comment (e.g., negative tone) as it may infringe the rights of the users to disagree. For example, for an offensive comment such as \textit{Why the f***ing phantom downvote?}, the paraphrased suggestions might be \textit{Why the pointless phantom downvote?} or \textit{Why the useless phantom downvote?} or \textit{Why the same phantom downvote?}. The user can then select one of options presented. Techniques used for paraphrasing are described in Section~\ref{subsection_offence_rephrasing}.

        The next section describes the OLRS-SE pipeline that implements the OLRS-SE framework (in Figure~\ref{fig:OLRS-SE_framework}) to reduce offensive language in SE communities.

    \subsubsection{OLRS-SE pipeline}         
    \label{subsection:OLRS-SE_pipeline}

\begin{figure}[h!]
            \includegraphics[width=\linewidth]{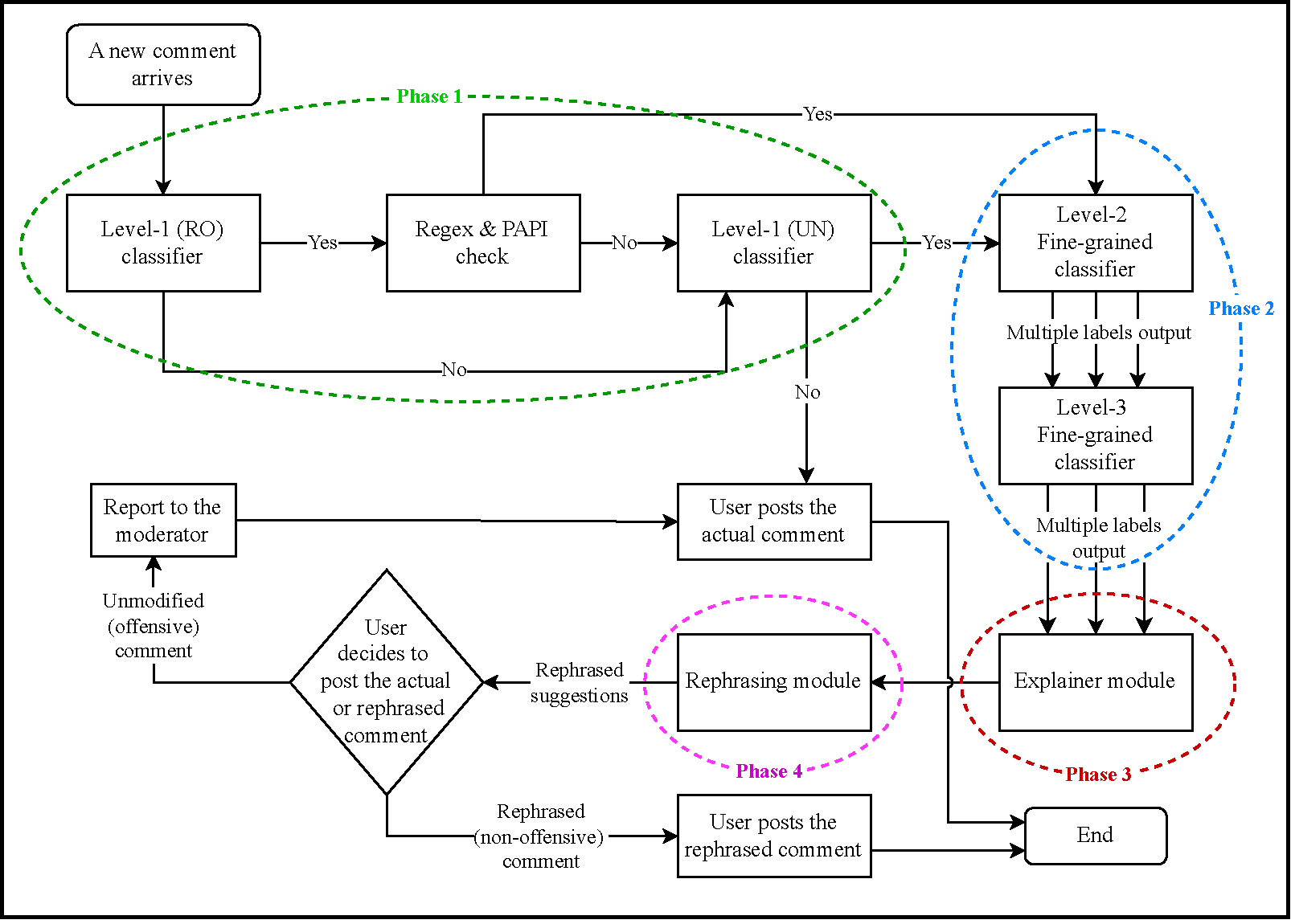}
            \caption{OLRS-SE pipeline}
            \label{fig:OLRS-SE_pipleline}
        \end{figure}

        We introduce a detailed pipeline that operationally combines the levels mentioned in Table~\ref{table:Fine_grained_taxonomy} and the framework presented in Figure~\ref{fig:OLRS-SE_framework} to implement the Offensive Language Reduction System. This pipeline of OLRS-SE operations is shown in Figure~\ref{fig:OLRS-SE_pipleline}. The four conceptual phases of the framework in Figure~\ref{fig:OLRS-SE_framework} are implemented through the four operational phases highlighted as dotted ellipses in Figure~\ref{fig:OLRS-SE_pipleline}.

        The first phase of the OLRS-SE pipeline (represented by the green ellipse) determines whether a newly constructed comment from a user is offensive. Here, we address the detection of RO and UN comments separately (i.e., we use separate classifiers for the two types of comments). Therefore, the first phase of the pipeline contains two offence detection classifiers: Level-1 RO and Level-1 UN classifiers. 
        
        When a new comment arrives, it passes through the Level-1 RO classifier which detects whether that comment is an RO offence. In order to detect the Level-1 RO class of offence, we developed twelve different ML models with different features (called L1.RO models) and selected the best, which are explained in Section~\ref{subsection:ML_model_building}. From the initial results reported  in~\citep{DBLP:conf/coin-ws/CheriyanSC20}, it was evident that a single ML model might not be enough to precisely detect all potential RO comments.  Therefore, we included a module that uses an additional rule-based filtering approach for detecting offence using regular expressions (Regex) and Perspective API scores (PAPI) outcomes to decide whether a comment is an RO offence~\citep{10.1145/3463274.3463805}. For this purpose, we collected the Regexes used by the SO Heat Detector bot\footnote{SO Heat Detector: \url{https://github.com/SOBotics/HeatDetector}}, and the complete Regex list is available online\footnote{Regex list: \url{https://github.com/SOBotics/HeatDetector/tree/master/domain/stackoverflow}}. PAPI\footnote{Perspective API: \url{https://www.perspectiveapi.com/}} offers a
        `toxicity’ score for a sentence in the interval of 0 (least toxic) and 1 (extreme toxic). Using the Regex and PAPI modules together has been found to be effective at detecting RO comments~\citep{10.1145/3463274.3463805}. If the Level-1 RO classifier predicts that a comment is offensive, it also needs to be positively identified by the Regex and PAPI module as an offence. Thus, RO detection is a two-step process for ensuring the robustness of our results. If the Regex and PAPI module confirms that the comment is indeed an RO offence, that comment goes through to the next levels of the pipeline for fine-grained offence detection. If the outcome of the Regex and PAPI module is negative, then the comment is not an RO offence.

        Even if the outcome of the Level-1 RO classifier is negative, that does not mean that the comment is non-offensive; rather, it can be mildly offensive. Therefore, that comment goes through the UN classifier which checks whether it is a mild offence or not. The same also happens when the Regex and PAPI module judges a comment to be non-RO outcome even though the comment had been identified as RO by the Level-1 RO classifier. In both these cases, the Level-1 UN classifier tests this comment to check whether it is a UN offence or not. If the outcome is positive, the comment goes to the next steps in the pipeline for processing. Otherwise, it is inferred that the comment is not an offensive one and this can be allowed to be posted online. For Level-1 UN detection, again, we developed a set of 12 different ML models called the L1.UN models (see Section~\ref{subsection:ML_model_building}) and used the best performing model for the development of OLRS-SE.

        The second phase of the OLRS-SE pipeline (represented by the blue ellipse) accommodates Level-2 and Level-3 classifiers of the taxonomy. In Level-2, a comment is classified based on the target of offence. In order to avoid bias owing to the highly unbalanced nature of the dataset, we split the model building process into two by decomposing the training dataset~\citep{DBLP:conf/ecml/VilaltaR03, DBLP:journals/nca/ElyanG16}. Therefore, we developed two different sets of twelve ML models for classifying Level-2 classes of RO offence. The first set of Level-2 models (L2.RO.1 models) classifies the \textit{Individual}, \textit{Generalized} and \textit{Others} classes while the second set of another 12 models (L2.RO.2 models) identifies the rest of the classes of the RO offence (\emph{Entity-abuse} and \emph{Self-abuse}). Similarly, we developed a set of 12 models to detect the \textit{Individual}, \textit{Generalized} and \textit{Others} classes of UN offence (L2.UN models). For the rest of the Level-2 UN classes, we did not develop models due to the scarcity of data points in our dataset.

        The Level-3 classifier identifies whether the comment possesses racial or swearing terms.  Similar to the previous levels, we developed another set of twelve models for the detection of swearing and racial classes (L3.RO models). In total, to identify classes corresponding to Levels 1 to 3, we developed a total of six sets of 12 models, a total of 72 models (6 * 12 = 72). These six sets of models developed for six specific tasks mentioned above are L1.RO, L1.UN, L2.RO.1, L2.RO.2, L2.UN and L3.RO.
    
        After Level-3 classifications, the comment proceeds to the explanation module, which is the third phase of the OLRS-SE pipeline (represented by the red ellipse). The explanation module is able to determine and highlight the offensive contents of the comment to inform the writer about the nature of the offence (offence type, target, and racial/swearing). The goal of the explanation module is to educate the user about the offensive nature of their post. This explanation may help them to modify their own post to a less or non-offensive one. For explaining the offence, we used the results from Level-1 ML models. This is explained in detail in Section~\ref{subsection_offence_rephrasing}. After the explanation module, the fourth and final phase in the OLRS-SE pipeline (represented by the pink ellipse) is the paraphrasing module that paraphrases offensive comments into less offensive or non-offensive ones. For paraphrasing, we developed two deep learning based large language models, and these are  also presented in Section~\ref{subsection_offence_rephrasing}.
    
        Even after the writers are informed about the nature of the offence, they may not want to reduce the toxicity and may still want to post the comment. They must be given the chance to do that. Otherwise, it would be construed as an infringement of their freedom of speech. Therefore, to further nudge the user to reduce offence, the user will be provided with the chance to choose amongst the original comment and a set of detoxified suggestions based on the original comment. If the user proceeds with their original comment, which is an offensive one, the moderators will be informed that the user has chosen to post an offensive comment (and this comment may not be publicly posted on the site by default). Otherwise, the writers can choose a detoxified comment from the provided suggestions and this post can be made available to the public.

    \subsection{Offence detection and classification (RQ2.1)}
    \label{subsection:ML_model_building}
    
        The first building block for the OLRS-SE is to successfully detect offensive language. We pose \textbf{RQ2.1} to address this: \textbf{\textit{How can we accurately detect and classify offensive language in online SE communities, using machine learning techniques}?} Prior work reported in~\citep{10.1145/3463274.3463805} presents a rule-based system for the same purpose. However, we use the  ML models to tackle this question for their potential to yield superior performance. We created both traditional and deep learning based ML models to train classifiers. 
    
        \subsubsection{Text preprocessing}
            For the data to be used by traditional ML models such as Random Forests (RF)~\citep{breiman2001random} and Support Vector Machine (SVM)~\citep{vapnik1995t}, we performed standard preprocessing techniques such as removal of usernames, non-alphabetic characters (i.e., numbers, punctuation marks and special characters) on all comments. For deep learning models, we just performed username removal. This is because removing stopwords (pronouns, prepositions and conjunctions) have been shown not to  have much effect on the performance of the deep learning models~\citep{DBLP:journals/corr/abs-1904-07531}. 
            
        \subsubsection{ML model building}  
        \label{subssection:ML_modelling}
            \textbf{Traditional ML models}: We built RF and SVM  models using a combination of three sets of features:  TF-IDF~\citep{ref1}, sentiment~\citep{DBLP:conf/icwsm/HuttoG14} and politeness~\citep{DBLP:conf/acl/Danescu-Niculescu-MizilSJLP13}. A set of five RF models and five SVM models with different combination of features were developed. We chose these two algorithms because prior work has demonstrated that for abuse detection in short text such as social media comments, RF and SVM are good candidate models~\citep{savarimuthu2020using}. 
            
            \emph{TF-IDF features} have been used by the Heat Detector bot of SO to detect offensive language where it is a part of the Naïve Bayes Multinomial model\footnote{TF-IDF used by Heat Detector bot: \url{https://stackapps.com/questions/7001}}. The classifier in the bot uses TF-IDF, along with an NGramTokenizer (min 1, max 3). We chose TF-IDF features in our work due to its demonstrated importance in prior work~\citep{savarimuthu2020using}.
    
            \emph{Sentiment features} play a major role in the decision-making process of answering questions and commenting in SO~\citep{10.1145/2661685.2661689, DBLP:journals/infsof/CalefatoLN18}. For example, questions exhibiting negative sentiments reduce the probability of getting successful answers or valuable comments from the SO community~\citep{DBLP:journals/infsof/CalefatoLN18}. Sentiment expresses the polarity of the language (neutral, positive and negative). Offensive or unfriendly language usually exhibits negative sentiment~\citep{schmidt_survey_2017}. So, we considered sentiment as a feature in our work. To analyse the sentiment, we used the VADER Python library\footnote{VADER package: \url{https://pypi.org/project/vaderSentiment}}. The sentiment-related features used in our work are sentiment polarity (neutral, positive or negative) and polarity score, which indicates how strongly the sentiment has been expressed. 
    
            The \emph{politeness features}\footnote{Politeness package: \url{https://cran.r-project.org/web/packages/politeness/vignettes/politeness.html}} are a set of 36 syntactic and social markers in natural language (expressing, e.g., gratitude, apology, acknowledgment and command). These features are used to evaluate politeness expressed in social settings such as expressing positive or negative politeness. Positive politeness involves expressing gratitude while communicating, but negative politeness often conveys ideas by using hedges (e.g., may be, might be) or informal titles (e.g., buddy, dude). These features are extensively used for politeness identification~\citep{YEOMANS2020131}. We used an R program to collect the politeness features of comments using the \textit{Politeness} package. Having identified the three sets of features, we implemented the ML models using Python's \textit{Scikit-learn} package with hyper-parameter tuned settings. We developed a total of 60 traditional models (10 models for each of the six tasks discussed in Section~\ref{subsection:taxonomy}).  Of these 10 models, there were 5 RF and 5 SVM models that employed different combinations of three features (see Figure~\ref{fig:Cross validated results} legend for the combinations). All the models were evaluated using 10-fold cross validation approach.
    
            \textbf{Deep learning models}: We developed deep learning models using the transformer-based language model BERT. To build the BERT models, we fine-tuned the pretrained BERT\textsubscript{Base-Uncased} and BERT\textsubscript{Base-Cased} versions using the  Huggingface Transformers Python package~\citep{wolf-etal-2020-transformers}, with the developer-recommended parameters: batch size: 16, maximum token length: 250 and dropout probability: 0.1. All comments were padded with the developer-recommended special tokens of \textit{[CLS]} and \textit{[SEP]}. We used a low learning rate ($3\times 10\textsuperscript{-5}$) and Adam optimizer with a base learning rate of $1\times 10\textsuperscript{-5}$ for training. The models were trained for three epochs to avoid overfitting. A total of 12 models were developed corresponding to the six tasks (6 * 2 = 12). 

            Overall, we created six sets of twelve different  models (6 * 12 = 72 models) for the detection and classification of offensive language. To summarize, for detecting the Level-1 classes of the taxonomy (RO and UN), we used two sets of ML models (L1.RO and L1.UN).  Similarly, to identify the targets of abuse (Level-2), we created three sets of models: two for RO target identification (L2.RO.1 and L2.RO.2) and one for UN target identification (L2.UN). Finally, for specific offence detection (Level-3), we used another set of models (L3.RO). Amongst the twelve different models in each set, there are two variants of BERT, five RF and five SVM models which were developed using different combinations of above-mentioned features (TF-IDF, Sentiment and Politeness).

    \subsection{Offence explanation and rephrasing (RQ2.2)}
    \label{subsection_offence_rephrasing}

\begin{figure}[ht]
            \includegraphics[width=\linewidth]{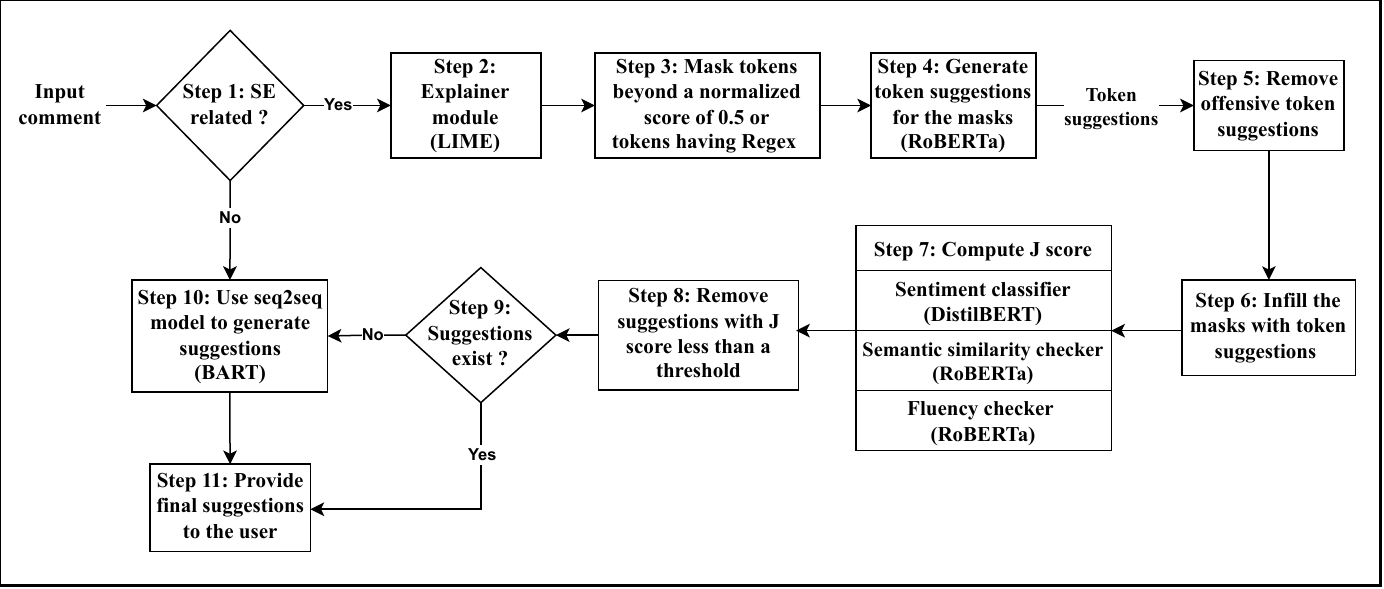}
            \caption{Paraphrasing module pipeline}
            \label{fig:Chapter9_paraphrasing}
        \end{figure}

     In this sub-section, we present the paraphrasing module which encompasses two aspects of the OLRS-SE framework (offence \textit{explanation} and \textit{rephrasing}), addressing RQ2.2: \textbf{\textit{How can we develop approaches to explain offensive language to its poster and also rephrase offensive ones into less offensive ones using  language models?}}. The objective of the paraphrasing module is to detoxify the offensive contents in textual comments and transform them into less offensive or non-offensive ones. To detoxify user generated comments, we propose a paraphrasing framework called Paraphraser, and the steps involved in paraphrasing a comment are shown in Figure~\ref{fig:Chapter9_paraphrasing}. Paraphraser adopts and modifies the Generate-Prune-Select (GPS) approach proposed by Zhu and Bhat~\citep{zhu-bhat-2021-generate}, along with the offence explanation methods to detoxify offensive comments. Through detoxification, the toxic comments are transformed into less offensive comments, which can then be recommended to the posters, from which the posters can choose the replacement comment.   

    The paraphraser module is implemented as 11 steps, as shown in Figure~\ref{fig:Chapter9_paraphrasing}. \emph{Step 1} involves checking whether the offensive comment is related to Software Engineering (SE). To achieve this, following Rahmi et al.~\citep{10.1145/3457784.3457796}, we built a corpus of SE words by collecting the most commonly occurring 5000 terms of SEthesaurus~\citep{DBLP:journals/tse/ChenCZX21}. If the comment is SE-related, toning-down the offensiveness of the comment preserves the disagreement nature and SE-relevance of the comment by keeping the negative sentiment and preserving the SE-related words of the comment (which is achieved through steps 2 to 8). \emph{Step 2} involves the identification of terms that are offensive.  For that, we used a LIME-based explanation module that returns a list of the most prominent words that contributed towards offensiveness. LIME~\citep{10.1145/2939672.2939778} is a widely used model-agnostic and post-hoc explanation method that locally explains ML model decisions.  LIME works with perturbed instances of inputs (i.e., comments). We set the number of perturbed instances of inputs to 500. Also, we set the number of features as 5, meaning that LIME identifies the most promising five words that caused offence. We limited the number of words to 5 since it was rare (in our experience) to find an offensive statement in the SO dataset with more than five offensive terms. For ML explanations using LIME, we used the \textit{Lime} package\footnote{LIME: \url{https://github.com/marcotcr/lime}} of Python. For a particular comment, the package returns a list of words in the comment and their corresponding scores as the explanation for a particular label (e.g., offensive label). The scores are the proportional probability contribution of a term towards the comment being assigned to a particular label. For example, for a particular offensive comment \textit{you and your stupid java answer}, LIME returns  the following list with the following top-5 terms: (stupid: 0.24732, your: 0.01414, you: 0.00514, answer: 0.00218, java: 0.00109). Since these scores are not normalized, we scale them to a range of 0 to 1 so that we are able to identify the most prominent tokens towards offensiveness (\emph{Step 3}). Moreover, by setting a threshold, we were able to remove potential polite words that LIME may propose as impolite~\cite{Wattanakriengkrai_2022, Yu_2024}. After the scaling process, the above-mentioned output of LIME  will be: (stupid: 1, your: 0.05299, you: 0.01644, answer: 0.00442, java: 0). Following similar work~\citep{Wattanakriengkrai_2022, Yu_2024}, we empirically adopted the LIME threshold as 0.5, where setting the value higher missed identifying terms that were offensive and setting the value lower resulted in including non-offensive terms. Tokens (i.e., terms) with scores higher than the threshold are selected for masking and the rest of the terms (potential impolite terms) are kept intact without masking. In addition, we also identify offensive terms using Regex patterns that need to be masked so that we do not miss any offensive terms.  
    In \emph{Step 4}, we generate a set of candidate paraphrases which can be used as substitutes for the masked terms (i.e., offensive terms) identified in step 3. We used RoBERTa~\citep{DBLP:journals/corr/abs-1907-11692}, a Masked Language Model (MLM), which is able to generate alternate tokens for a special class of token called \textit{[mask]} tokens identified in step 3. While RoBERTa is able to generate a large number of alternatives for a masked word, in \emph{Step 5} we limit our results to the top five non-offensive token suggestions for each offensive token. 
    
    Then, in \emph{Step 6} we fill the masks in a sentence with rephrased words. In \emph{Step 7} we compute the J score metric~\citep{Krishna2020ReformulatingUS} which incorporates three aspects: style, context and grammatical accuracy for the evaluation of the generated paraphrases. In \emph{Step 8} we remove suggestions (rephrased comments) with J score less than 0.8\footnote{This was adopted empirically during pilot experimentation, where lower thresholds admitted a noticeably larger number of semantically unrelated candidates, while higher thresholds were observed to exclude relevant matches. Our objective was therefore to adopt a conservative threshold that retained highly similar candidates.}.  Then we select top-3 suggestions from this set and present them as recommendations to the writer. 
    If the comment is not related to SE (as a result of Step 1), or if there is no paraphrased comment meets the J score threshold (\emph{Step 9)} then a seq2seq model that has been developed is used. We employed the counter-narrative approach in~\citep{chung-etal-2019-conan} to develop this seq2seq model. Thus, the objective of the seq2seq model developed using BART~\citep{lewis-etal-2020-bart} is to generate counter-narratives for non-SE-related offensive comments (\emph{Step 10}). To achieve this, we developed a curated dataset of comments that contain offensive language, along with alternative text for these comments~\citep{Mukherjee_2023}. The dataset was prepared in such a way that for a source sentence (an offensive comment), there is a target sentence (non-offensive suggestion). The dataset contains 500 RO comments as source sentences that are offensive and not related to SE discussions. The target set contains 500 sentences that are non-offensive comments. A seq2seq model trained on this dataset produces an alternative sentence to an offensive sentence that does not have offensive language. This counter sentence, when presented as an alternative to the offensive comment, could be chosen by the authors and thus enable a more civilized conversation. For example, a comment such as \textit{f*** [that] question [was] allowed to stay open [but] mine get closed} can be rephrased as \textit{I am offended because my question got closed}. As mentioned above, we aim to preserve the negative tone of the offensive comment that expresses disappointment in the paraphrased suggestions, but with reduced offensiveness. The training data can be found online\footnote{Seq2seq model training data: \url{https://doi.org/10.6084/m9.figshare.19513705}}. 

    The exertion of J-score is limited to the suggestions generated by the modified GPS model only, but was not used for the seq2seq model. Given that the seq2seq model was trained on a parallel corpus (a dataset of generic offensive comments that are not related to SE domain and their corresponding general non-offensive suggestions), the surrounding context of the input comment (offensive) is not being preserved to create non-offensive output suggestions. In such a scenario, the  J-score calculation of the non-offensive suggestions is pointless as the style and context of the input comment and output suggestions vary significantly. A more detailed explanation of the 11 steps in Figure~\ref{fig:Chapter9_paraphrasing} can be found in ~\ref{Sec: Appendix Paraphraser Module}.    The next section presents the results of various ML models and the OLRS-SE system built for the detection and reduction of offensive language.

\section{Results}
\label{sec:results}

\begin{figure*}[ht!]  
        \centering
            \includegraphics[width=\linewidth]{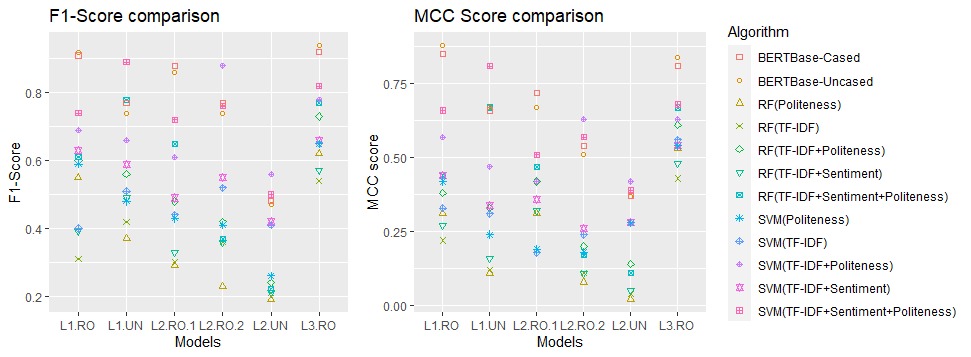}
            \caption{Cross validated results of ML models in OLRS-SE}
            \label{fig:Cross validated results}
        \end{figure*}

     Figure~\ref{fig:Cross validated results} presents the cross validated F1-score and MCC scores of six sets of twelve ML algorithms used to achieve six specific tasks of the OLRS-SE pipeline discussed in Section~\ref{subssection:ML_modelling}. The full results in tabular form is available online\footnote{ML model results: \url{https://doi.org/10.6084/m9.figshare.27139785}}. L1.RO models identify RO comments at Level-1. Twelve different algorithms were developed for this purpose. The other five tasks are identifying UN comments at Level-1 (L1.UN), identifying targets of RO comments in Level-2 (L2.RO.1 and L2.RO.2), identifying targets of UN comments in Level-2 (L2.UN), and identifying swearing and racial comments in Level-3 (L3.RO). Having developed these models, for each specific task, the ML models with the highest F1-score and MCC were selected to implement OLRS-SE. We report the results based on F1-score results because it is the harmonic mean of both precision and recall, and is an effective metric to evaluate the performance of class-imbalanced data model~\citep{DBLP:journals/tkde/HeG09, lipton2014optimal}. For class imbalanced data sets, MCC is preferred since it considers the balance across the four confusion matrix quadrants~\citep{chicco2020advantages}. Since the rank-ordered results for F1-score and MCC metrics for the algorithms are similar, we only report and discuss patterns of results using the F1-score below.

        \textbf{Offence detection and classification}: For Level-1 RO (L1.RO) detection, the BERT\textsubscript{Base-Uncased} has the highest F1-score and for Level-1 UN (L1.UN), the SVM model (with TF-IDF+Sentiment+Politeness features) outperformed all others. For the classification of Level-2 RO classes (Individual, Generalized and Others) (L2.RO.1), BERT\textsubscript{Base-Cased} surpassed others, and for the detection  of Entity abuse and Self-abuse (L2.RO.2), an SVM model with TF-IDF and Politeness features performed best. For the classification of Level-2 UN (L2.UN) classes, an SVM model with same features excelled than others. For the detection of Level-3 classes (L3.RO), BERT\textsubscript{Base-Uncased} offered the best results. Generally, for the detection of RO offence, the BERT models outperformed the traditional ML models (RF and SVM), and, for the detection of UN comments, the SVM models surpassed the rest. 

        It can be observed that the best performing models across all levels except L2.UN, achieve an F1-score of 0.88 or above. This shows that the best performing models are effective in classifying comments into different labels (at different levels). Only for L2.UN, the top result has a F1-score of 0.56. This we believe is because of the nuances in identifying the mildly offensive comments which we elaborate in the discussion section.

        \begin{figure}[h!]    
            \centering
            \includegraphics[width=\linewidth]{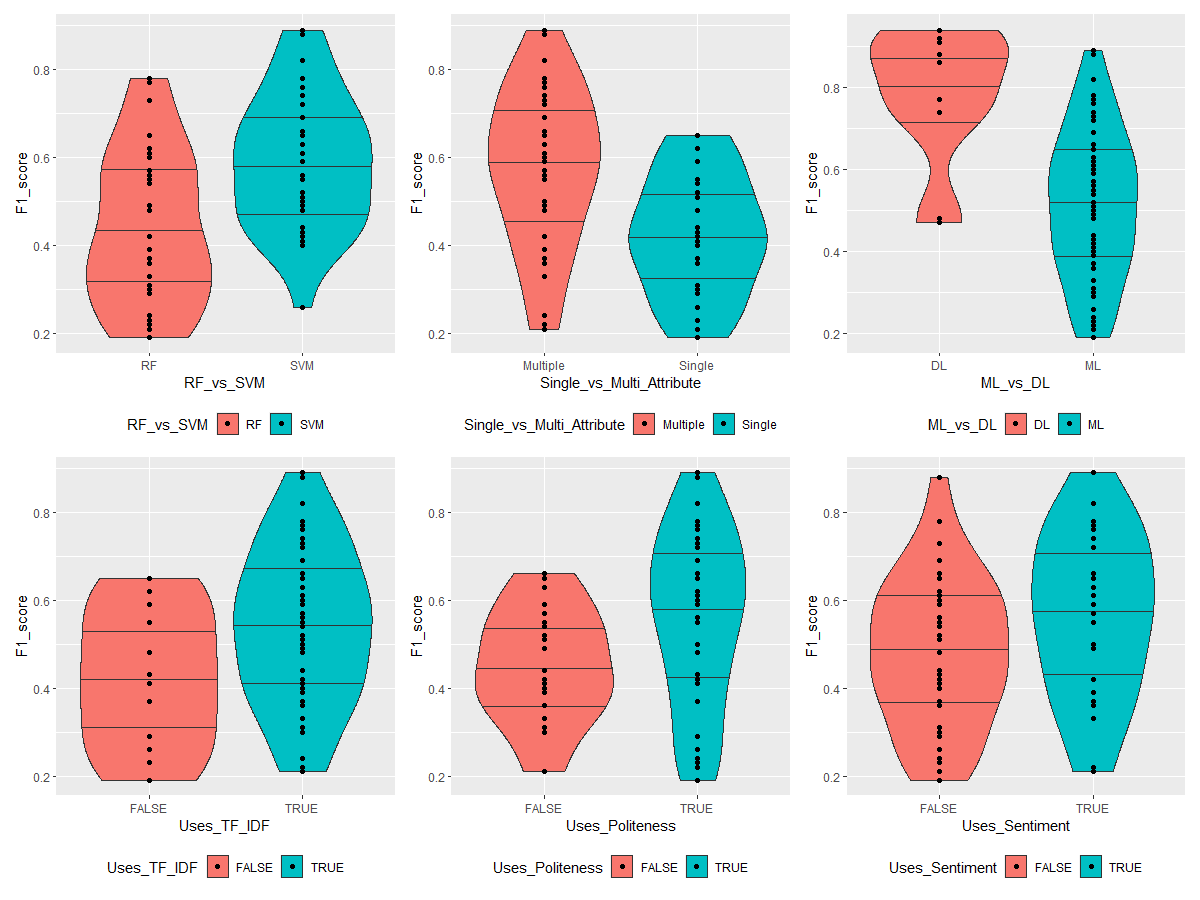}
            \caption{Comparison of results for six variable groups}
            \label{fig:group_comparison_results}
        \end{figure}

        We also analysed results obtained from the models, by assigning each model into one of two groups based on six different grouped variables. The visual comparison of F1-score results for these six grouped-comparisons are shown as violin plots in Figure \ref{fig:group_comparison_results}. The first variable considered was whether a model uses RF or SVM algorithms. All results from RF models were assigned to one group (5 * 6 = 30 results)  and the results from SVM models were assigned to the other group (5 * 6 = 30 results).  The top-left result in Figure \ref{fig:group_comparison_results} shows the distribution of F1-scores for RF and SVM models (in red and blue respectively). It can be observed that the median score for SVM models was higher than RF models (see the middle horizontal line in a violin plot). The Mann-Whitney U-test result for these two groups showed that there is a statistically significant difference between the F1-scores obtained from RF and SVM models. SVM models had higher F1-scores than RF models (p $<$ 0.05). Second, we investigated if there was a significant difference between traditional ML models that considered just one feature or attribute such as TF-IDF (i.e., 4 * 6 = 24 models) vs. those that considered multiple attributes (6 * 6 = 36 models). The Mann-Whitney U-test results shows that the multi-attribute models produced statistically higher F1-scores than single attribute models (p $<$ 0.05). Third, we investigated the difference between traditional ML (the ten variants involving RF and SVM, i.e., 10 * 6 = 60) vs. DL  (2 * 6 = 12) models. The Mann-Whitney U-test result showed that DL algorithms had significantly higher F1-scores than ML algorithms (p $<$ 0.05).  Fourth, there was no statistically significant difference in the group that use TF-IDF attribute (8 * 6 = 48 models) and the group that did not use TF-IDF variable (2 * 6 = 12 models). Fifth, we observed that among traditional ML models using politeness variables (6 * 6 = 36 models) yielded statistically higher F1-scores than those (4 * 6 = 24 models) that did not employ these variables (p $<$ 0.05). Sixth, the difference in F1-scores in traditional models that use sentiment variables (4 * 6 = 24 models) vs. the ones that did not use the sentiment variables (6 * 6 = 36 models) was not statistically significant.  
        
        These results in highlight the promise of using DL algorithms over traditional ML algorithms (DL vs. ML), particularly if just one family of algorithms must be selected for all classification tasks. Also, if traditional ML algorithms were to be used for classification tasks, our results underscore the importance of politeness variables in detecting offence.

        \textbf{Offence explanation and rephrasing}: To evaluate the performance of the rephrasing module, we sampled 200 SE-related comments from the SO sample rude/offensive dataset\footnote{SO RO comments: \url{https://meta.stackoverflow.com/questions/326494}} using the SE relatedness check. We ensured that these comments are highly offensive (RO), using the OLRS-SE Level-1 RO classifier that includes the Regex and PAPI check. These highly offensive comments were paraphrased to less offensive or non-offensive comments as discussed in Section~\ref{subsection_offence_rephrasing}. The paraphraser module generated sentences with an  average J score of 0.8 (out of 1) indicating that our system works reasonably well in paraphrasing offensive comments. The examples in Table~\ref{Table:GPS_sample_results} shows three sample results from our system. We should note that, for 18 SE-related comments (9\%), the paraphrasing is generated by the seq2seq model since the candidate suggestions from the modified GPS model were unable to meet the J score threshold. In other words, 91\% of the comments were paraphrased using the GPS rephrasing module and 9\% using the seq2seq model. These 200 comments and their paraphrases, along with corresponding evaluation metrics, can be found online\footnote{SE-related offensive comments: \url{https://doi.org/10.6084/m9.figshare.19860097}}. 

            \begin{table}[ht]           
            \begin{tabular}{|c|c|}
                \hline
                \textbf{Input comment} & \textbf{Paraphrased suggestions with J score in brackets} \\  \hline
                
                \multirow{3}{4.25cm} {\textit{I am sucked with microsoft}} 
                        & I am struggling with microsoft (0.83629036) \\\cline{2-2}
                        & I am stuck with microsoft (0.8240543) \\\cline{2-2}
                        & I am done with microsoft (0.68389195)  \\\cline{1-2}
                             
                \multirow{3}{4.25cm} {\textit{Its a hacking tool, delete the fucking post}}
                        & Its a hacking tool, delete the damn post (0.987) \\\cline{2-2}
                        & Its a hacking tool, delete the offending post (0.948) \\\cline{2-2}
                        & Its a hacking tool, delete the entire post (0.948) \\\cline{1-2}

                \multirow{3}{4.25cm} {\textit{Do you have any stupid specific programming question?}}
                     & Do you have any other specific programming question? (0.912) \\\cline{2-2}
                     & Do you have any very specific programming question? (0.911) \\\cline{2-2}
                     & Do you have any more specific programming question? (0.888) \\\cline{1-2}
                
                \end{tabular}    
                \caption{Modified GPS sample output}
            \label{Table:GPS_sample_results}
            \end{table}

        \textbf{Reliability results}: It should be noted that we developed the OLRS-SE using the data from the SO dataset. In order to demonstrate the scalability of our study, we tested comments obtained from three other SE communities (GitHub, Gitter and Slack). Based on data from previous work of Cheriyan et al.~\citep{10.1145/3463274.3463805}, and using purposive sampling, we collected 509, 511 and 508 comments from GitHub, Gitter and Slack datasets, respectively, to be used as test datasets. These samples contained roughly equal number of offensive and non-offensive comments. Table~\ref{table:results_1} presents the performance of the best models used in OLRS-SE to detect the offence classes in three other SE communities. The accuracy of offence detection and classification varied from 57\% to 100\%. However, comparatively lower F1-scores and MCC scores (compared to results for the SO dataset) indicate the challenges with models trained using a particular dataset (SO dataset) to precisely detect and classify the offence classes of another dataset (from a different platform -- e.g., Gitter). Moreover, varying comment length and vocabulary differences might have contributed towards lower scores presented in Table~\ref{table:results_1}. One approach is to fine-tune models on the new dataset for better results. Alternatively, generalized models can be developed from scratch using input data from multiple sources (i.e., combining results from all four datasets).

      \begin{table}[h!]        
            \centering
            \begin{tabular}{|p{2.5cm}|c|c|c|c|c|c|}     
            \hline
            \textbf{Label} & \textbf{Platform} & \textbf{Accuracy}& \textbf{Precision} & \textbf{Recall} & \textbf{F1-Score} & \textbf{MCC}\\
            \hline       
           
    \multirow{3}{*}{L1 RO} & GitHub & 0.85 & 1.00 &	0.14 & 0.25 & 0.35 \\ \cline{2-7}
	 & Gitter	& 0.88 &	1.00 &	0.38 &	0.55	& 0.58 \\ \cline{2-7}
	 & Slack	& 0.92 &	1.00 &	0.30 &	0.46 &	0.52 \\ \hline 
   
    \multirow{3}{*}{L1 UN}	& GitHub &	0.61 &	0.25 &	0.31 &	0.28 &	0.01 \\\cline {2-7}
	 & Gitter & 	0.59 & 	0.25 &	0.60  &	0.36 &	0.15 \\\cline {2-7}
      & Slack &	0.57 &	0.27 & 0.58 &	0.37 &	0.12 \\ \hline
    
    \multirow{3}{*}{L2 Individual} &	GitHub &	0.80 &	0.45 &	0.47 &	0.46 &	0.34  \\\cline {2-7}
    	 & Gitter &	0.72 &	0.30 &	0.31 &	0.31 &	0.13 \\\cline {2-7}
    	& Slack	& 0.69 &	0.31 &	0.24 &	0.27 &	0.08 \\ \hline
    
    \multirow{3}{*}{L2 General} &	GitHub &	0.96	& 0.71 &	0.42 &	0.53 &	0.53 \\\cline {2-7}
    	& Gitter &	0.93 &	0.57	& 0.35 &	0.43 &	0.41 \\\cline {2-7}
    	& Slack &	0.97 &	0.17 &	0.22 &	0.19 &	0.18 \\ \hline
    
    \multirow{3}{*}{L2 Ent. abuse} & GitHub &	0.96 &	0.70 &	0.32 &	0.44 &	0.46 \\\cline {2-7}
    	& Gitter & 0.94 &	0.35 &	0.41 &	0.38 &	0.35 \\\cline {2-7}
    	& Slack &	0.95 &	0.25 &	0.11 &	0.15 &	0.14 \\ \hline
    
    \multirow{3}{*}{L2 Self-abuse}	& GitHub &	0.99 &	0.33 &	0.20 &	0.25 &	0.25 \\\cline {2-7}
    	& Gitter &	0.95 &	0.36 &	0.18 &	0.24 &	0.23 \\\cline {2-7}
    	& Slack &	0.94 &	0.13 &	0.04 &	0.06 &	0.05 \\ \hline
    
    \multirow{3}{*}{L2 Others} &	GitHub &	0.78 &	0.67 &	0.32 &	0.44 &	0.35 \\\cline {2-7}
    	& Gitter &	0.84 &	0.68 &	0.74 &	0.71 &	0.60 \\\cline {2-7}
    	& Slack &	0.86	& 0.64 &	0.63 &	0.63 &	0.54 \\ \hline
    
    \multirow{3}{*}{L3 Racial} &	GitHub &	0.99 &	1.00 &	0.25 &	0.40 &	0.50 \\\cline {2-7}
    	& Gitter &	1.00 &	0.80 &	0.80 &	0.80 &	0.80 \\\cline {2-7}
    	& Slack &	NA &	NA &	NA	& NA &	NA \\ \hline
    
    \multirow{3}{*}{L3 Swearing} &	GitHub	& 0.83 &	1.00 &	0.11 &	0.20 &	0.30 \\\cline {2-7}
    	& Gitter &	0.86 &	1.00 &	0.31 &	0.48 &	0.52 \\\cline {2-7}
    	& Slack &	0.91 &	1.00 &	0.25 &	0.41 &	0.48 \\ \hline

            \end{tabular}    
            \caption{Performance of Stack Overflow models on other SE datasets}
            \label{table:results_1}
        \end{table}

        \subsection{Additional validation}  
        \label{subsec:Addit_validation}

         \begin{table}[h]        
            
            \centering
              \begin{tabular}{|c|c|c|c|c|} \hline
            \textbf{Level} & \textbf{Label} & \textbf{Accuracy (\%)} & \textbf{ $\kappa$-score} & \textbf{Agreement level} \\\hline
            \multirow{2}{*} {Level-1}
                & RO  & 94 & 0.91 & Almost perfect \\\cline{2-5}
                & UN  & 86 & 0.82 & Almost perfect \\\hline                
            \multirow{9}{*} {Level-2}
                & RO Individual & 79 & 0.82 & Almost perfect \\\cline{2-5}
                & RO Generalized & 81 & 0.80 & Substantial\\\cline{2-5}
                & RO Entity abuse & 71 & 0.78 & Substantial\\\cline{2-5}
                & RO Self-abuse & 89 & 0.90 &  Almost perfect\\\cline{2-5}
                & RO Others & 76 & 0.71 & Substantial \\\cline{2-5}
                & UN Individual & 71 & 0.67 & Substantial\\\cline{2-5}
                & UN Generalized & 72 & 0.59 & Moderate\\\cline{2-5}
                & UN Entity abuse & 77 & 0.62 & Substantial\\\cline{2-5}
                & UN Others & 69 & 0.58 & Moderate\\\cline{1-5}
            \multirow{2}{*} {Level-3}
                & Racial & 87 & 0.91 & Almost perfect\\\cline{2-5}
                & Swearing & 89 & 0.82 & Almost perfect\\\cline{1-5}
            \end{tabular}       
            \caption{IRR results of offence classes in OLRS-SE}
            \label{Table:chapter8_Ml_models_results_validation}
        \end{table} 

        Having developed these models and evaluated their effectiveness using well-known metrics such as F1-score and MCC, we also conducted an additional round of human validation where we compared results from the models vs. human evaluation (with two participants).  We collected a new set of 100 comments each from SO and Gitter. SO comments were collected from the sample SO rude, offensive or unwelcoming comments available online\footnote{SO sample RO/UN comments: \url{https://meta.stackoverflow.com/questions/326494}} and an equal amount of comments from Gitter were collected from prior work of Parra et al.~\citep{Parra_MSR}.  We used the best performing models of OLRS-SE to evaluate the classification performance. The first author first read each comment in the sample and recorded whether they agreed with the output from the OLRS-SE system. Then, we recruited an external participant (a PhD candidate in computing) for validation who also followed the same procedure as the first author. Then, consensus between the two evaluators was measured using Cohen's Kappa metric (see column 4 of Table \ref{Table:chapter8_Ml_models_results_validation}). The average Kappa score was 0.76, which shows substantial agreement between the two evaluators, with scores for different tasks ranging from 0.58 to 0.91. We also obtained satisfactory results for the comparison between model prediction and the consensus-based results from the two evaluators as shown in column 3 of Table~\ref{Table:chapter8_Ml_models_results_validation}. The average accuracy was 81\%.  The comments and coding results can be found online\footnote{IRR file: \url{https://doi.org/10.6084/m9.figshare.20154794}\label{IRR_Footnote_Reference}}.

        For evaluating the performance of the rephrasing module, we used the set of SO and Gitter comments mentioned above. Among the set of 200 comments, only 38 needed paraphrasing since only those were both SE-related and highly toxic. An IRR check was performed to evaluate the quality of the top three paraphrased suggestions and obtained a Kappa score of 0.71 among the first author and an external participant. This shows substantial agreement on the independent coders. The comments and their coding can be found online\textsuperscript{\ref{IRR_Footnote_Reference}}. The coding rules can also be found online\footnote{OLRS-SE coding rules:~\url{https://doi.org/10.6084/m9.figshare.32852210}}. The screenshot of the OLRS-SE tool developed based on the research reported here, which integrates all four aspects: offence detection, classification, explanation, and rephrasing, can be found here\footnote{OLRS-SE tool screenshot: \url{https://doi.org/10.6084/m9.figshare.30488333}}.  In the next section we provide a discussion of our results and the threats to validity.

\section{User evaluation}
\label{sec:user_evluation}       

    To evaluate the perceived utility of the developed Offensive Language Reduction System (OLRS-SE), we developed a web-based system. The user interface of the system connects to the OLRS-SE pipeline described in Section~\ref{subsection:OLRS-SE_pipeline} and presents the results to the users. The tool solicits users to enter a comment up to 250 characters long, and it checks whether it is offensive. For evaluation purposes, we prompted the users to enter SE-related offensive comments so that the classification and paraphrasing of the system can be experienced and evaluated, rather than entering a non-offensive comment. We developed the system using Deepnote\footnote{Deepnote: \url{https://deepnote.com}} and deployed the tool using Anvil\footnote{ Anvil: \url{https://anvil.works}}. The functionalities of the tool have been demonstrated in a YouTube video\footnote{OLRS-SE demo video-1: \url{https://youtu.be/wDlr7TrtvL8}}. An additional video demonstrating more examples is also available online\footnote{OLRS-SE demo video-2: \url{https://youtu.be/5L0ct7RVc0U}}.

      \begin{figure}[h!]            
            \includegraphics[width=\linewidth]{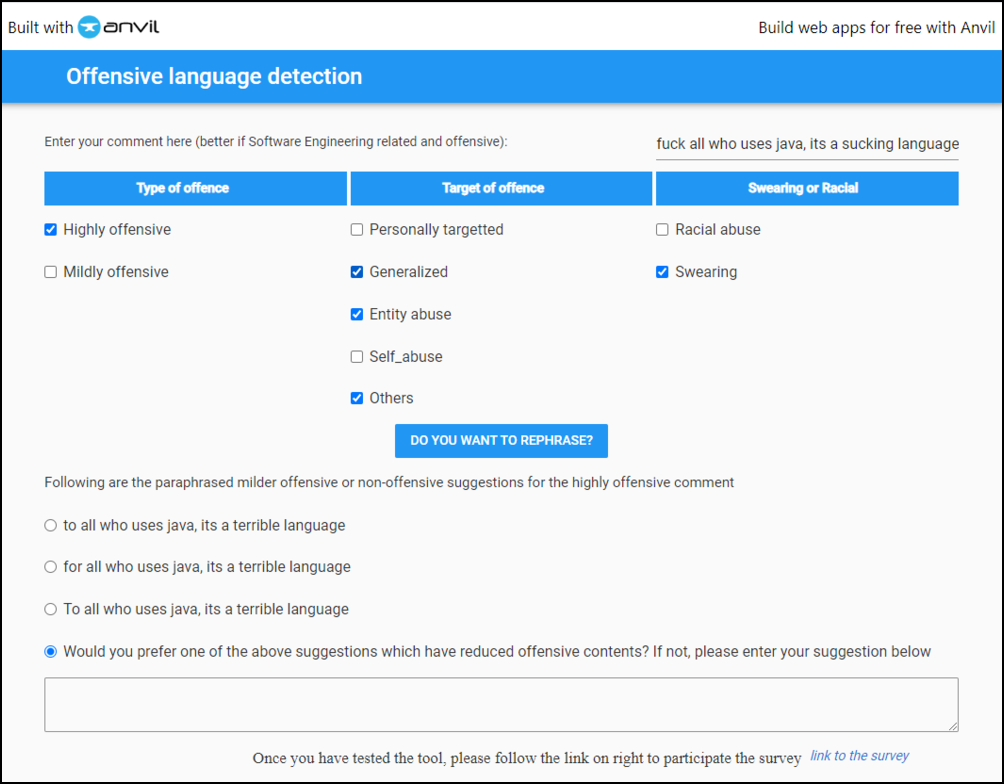}
            \caption{OLRS-SE tool sample screenshot}
            \label{fig:OLRS-SE results}
    \end{figure}  
        
    The tool presents three types of categorizations as shown in Figure~\ref{fig:OLRS-SE results}:  the type of offence, the target of offence and specific-fine grained classes. For an offensive comment, the tool can detect the classes of offence, as mentioned in the detailed taxonomy in Section~\ref{subsection:taxonomy}. To achieve these outputs, the tool uses the best performing ML models discussed in Section~\ref{subsection:ML_model_building}.
    
    The tool is also able to provide the user a set of rephrased suggestions for highly offensive comments, which will appear when the user presses the \textit{Do you want to rephrase?} button. The user may choose one of the recommended suggestions as the comment to be posted instead of the offensive comment. If they are not so inclined to choose one of the suggestions, they can also rephrase their own comment instead of posting an offensive comment, as shown in the last part of Figure~\ref{fig:OLRS-SE results} (see the option corresponding to the fourth radio button).

    \subsection{Methodology for user evaluation}
        To evaluate the developed system, we sought participants who studied or worked in computing-related areas. They were chosen because of their domain knowledge and the  potential to have prior experience of using online SE platforms. After obtaining ethical approval from our university, participants were recruited to evaluate the system. After consenting to participate, a task description document was given to the participants which detailed the tasks to be performed by them. Then, a demonstration of the developed web tool was presented to the participants, followed by the completion of three simple tasks by the user. The first two tasks were the detection and classification of RO and UN comments, respectively. For the third task, the user needed to submit their own comment to evaluate the detection, classification and rephrasing of an offensive comment. 

       Upon the completion of the three tasks, the participants were  asked to complete a questionnaire to evaluate the quality of the results returned by the system. The questionnaire comprised three parts: 1) background questions, 2) questions on the usefulness and usability of the system using selected questions from the System Usability Scale (SUS)~\citep{brooke1996quick} and the Technology Acceptance Model (TAM)~\citep{davis1985technology} questionnaires, and 3) specific-functionality related questions~\citep{Kitchenham2008}. The full questionnaire can be found online\footnote{OLRS-SE questionnaire: \url{https://doi.org/10.6084/m9.figshare.22689502}}, which comprised a total of 30 questions.

    \subsection{Results of user evaluation}   
    \label{subsection:user_evaluation_results}
\begin{figure}[ht]
             \setlength\abovecaptionskip{-30px}
            \includegraphics[width=\linewidth]{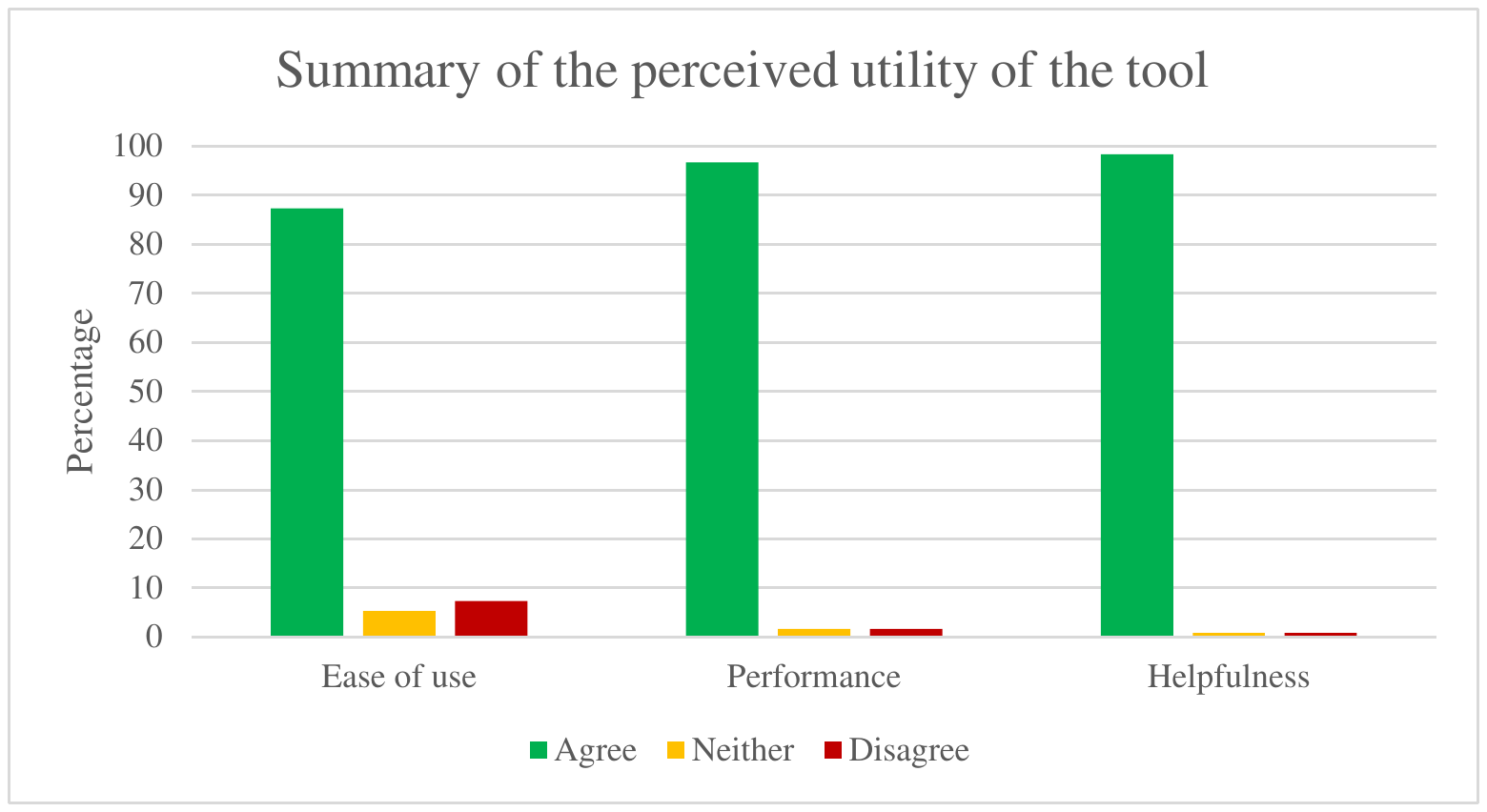}
            \caption{Summary of the specific evaluation of the tool}
            \label{fig:OLRS-SE_evaluation_Summary}
        \end{figure}

        We collected feedback from 30 participants with different backgrounds. Among the participants, 46\% were university students and the rest were computing professionals including university staff. 63\% of the participants were males and the rest were females.  We grouped the questions based on three themes: \emph{ease of use}, \emph{performance} and \emph{helpfulness} of the system. Figure~\ref{fig:OLRS-SE_evaluation_Summary} presents the summary of the results for these themes, which is the average of grouped responses. It shows that 87.3\% of the participants found that the tool is easy to use and 97\% of the participants were satisfied with the tool performance. Also, 98\% of them expressed that the tool is helpful to detect, classify and reduce offensive language in SE communities. Detailed user evaluation details\footnote{User evaluation data: \url{http://doi.org/10.6084/m9.figshare.23206373}} and theme-based aggregation details\footnote{Aggregated user evaluation data: \url{http://doi.org/10.6084/m9.figshare.23292416}} can be found online.        

        In order to solicit descriptive suggestions from the participants, we used two questions at the end of the survey on aspects they liked in the system and what they would like to be improved. Regarding the functionalities they found useful, one wrote ``\textit{The ability to adjust swearing without removing good sentiment was nice}''. This supports our attempts to maintain the overall tone of the statement (e.g., keeping user's sentiments while rephrasing highly offensive phrases). Another noted ``\textit{I find categorizing the offensive comments significant as it gives an idea on the sort of areas an offensive comment target [sic]}”. One suggestion for future work was to create a plugin that has the features of the OLRS-SE as expressed by a user who noted ``\textit{The OLRS-SE system is useful to detect offensive language and also very useful to get paraphrased comments. I would like to see and use when this [is] integrated to any other online platforms that are currently used or even to get as a google extension”}. A few opined that the system needs improvement, especially for improving the speed of providing paraphrased suggestions. We believe deploying the models in faster servers can address this issue.

        While expressing feedback on other ways of reducing offensive language in SE communities that are not yet a part of the tool, making ML models to continuously learn from user feedback was suggested. The feedback noted ``\textit{If there is a way you can add continuous online learning and improvement such that if the user is not satisfied with the feedback (rephrased comments), he can make suggestions and the ML model will improve from user feedback as well}''. This is a useful suggestion for future work where human-in-the-loop is incorporated as a part of continuous learning, and this can be achieved when the tool is integrated as a part of real-life systems such as Stack Overflow. In the next section we provide a discussion of our results and the threats to validity. 

\section{Discussion}
\label{sec:discussion}
        Our goal was to investigate the nature and prevalence of offensive language in SE communities and also propose an approach to reduce them. We discuss our findings for the two research questions posed.

        \textbf{Nature and prevalence of offensive language}: RQ1 focussed on the nature of offensive comments and their prevalence. Our work extended the preliminary taxonomy presented in~\citep{DBLP:conf/coin-ws/CheriyanSC20}, and has presented a more comprehensive taxonomy which comprises three levels as shown in Table~\ref{table:Fine_grained_taxonomy}. The taxonomy identifies the level of offence -- i.e., RO or UN (Level-1), detects the targets of offence (Level-2) and detects the specific fine-grained classes of  offence (Level-3).

        Table~\ref{table:Fine_grained_taxonomy} also presented the prevalence of these comments in SO. These results show that offensive language is indeed present in online SE communities, and we need approaches and systems that can help reduce such comments at source (i.e., when they are posted initially on a platform like SO). Also, we found that offensive language is an issue in other platforms (GitHub, Gitter and Slack) too. Our results here adds to growing voices of other researchers in this domain~\citep{Sarker_2023, Ferreira_2022, DBLP:conf/kbse/Sultana22a, DBLP:conf/enase/Sayago-HerediaC22, Sarker_2022} that offensive language use must be identified and removed.

        \textbf{A system to detect and reduce offensive language}: RQ2 focussed on reducing offensive language through a software system called OLRS-SE that detects, classifies and rephrases  offensive language using ML techniques, and thus helps reduce offensive language. To achieve the OLRS-SE framework that labels comments using the detailed taxonomy, we proposed an OLRS-SE pipeline (see Figure~\ref{fig:OLRS-SE_pipleline}). The pipeline is an implementation which operationalizes the detection of the  three levels of the taxonomy, and finally paraphrases the highly offensive comments to less offensive ones or non-offensive ones, using text detoxification techniques. The first phase of the OLRS-SE pipeline was the detection of offensive language. That included the detection of RO and UN classes of offences.

        \textbf{Offence detection and classification}: Sub-question {RQ2.1} involved developing and evaluating the performance of six sets of ML models (12 in each set) targeting specific classification tasks (e.g., identifying RO and UN classes) using the SO dataset. We found that ML models varied in their performance.  For example, at Level-1, the BERT-based models provided very good results for RO detection (F1-score of 0.92). For UN detection on the hand, the best model was SVM. Of the six sets of models developed as reported in Figure~\ref{fig:Cross validated results}, in three of them, the best performing model based on F1-score used BERT (L1.RO, L2.RO.1, and L3.RO), and in the other three, SVM models that included the politeness feature as one of the features fared the best (L1.UN, L2.RO.2, L2.UN). Our statistical analysis shows that deep learning models produced statistically better results than traditional ML models across all the six classification tasks.

        The reason why the BERT model wasn't able to perform as well as the SVM model on the UN dataset is because, the deep learning models were unable to identify or generalize the nuanced textual features contributing towards the unwelcoming contents, possibly owing to the use of a comparatively smaller dataset for training ($<$10000). Also, SVM model used explicit politeness features that help distinguish polite comments from impolite ones. In contrast, in RO dataset which had large amounts of data, BERT was able to perform better than the SVM model. Additionally, the best result for L2.UN was not high (0.56), which was produced by SVM. Other traditional ML model (RF) also fared poorly on this dataset. We believe this is because of the complex nature of identifying the problem (i.e., assigning the target of a mildly offensive comment). This is also the reason why the BERT models struggled to achieve good results in this task (F1-score of 0.47 and 0.48 respectively). We also noticed that for traditional algorithms (RF and SVM) that considered a single feature (e.g., TF-IDF) fared relatively poorly than multi-attribute feature set (e.g., TF-IDF and politeness). All top-performing traditional models (SVM-based models) included politeness features. Even in the three cases where BERT models was better, the SVM models with TF-IDF, Sentiment and Politeness fared closer to these models (i.e., these models included the politeness feature). This underscores the role of politeness features in the success of these models. Our statistical analysis also highlights the significance of the politeness features. Our main contributions here are a) the proposal of a fine-grained taxonomy for offence classification with three levels and b) the operationalization of the taxonomy through automated detection and classification of offences using ML approaches. 

        The reported cross-platform analysis was done as a pilot study on other datasets. However,  we observed significant vocabulary differences and varying length in the comments of other platforms. For example, GitHub comments found to be long and slack comments were the short in general. Gitter comments were of similar length as that of SO, which produced similar result metrics to SO comments. Given that we limited our comment length to 250 characters, this might have affected the performance of ML models on various classes of offence detection in GitHub comments. 
        
        \textbf{Offence explanation and paraphrasing}: In addressing RQ2.2, we developed solutions to reduce offensive content in textual comments using paraphrasing approaches. There were two main pathways for paraphrasing an offensive comment: a modified GPS pipeline to paraphrase SE-related comments and a counter-narrative generation approach for non-SE comments. The paraphrasing module consumes input from the LIME-based explainer module to locate the most offensive tokens in a comment. We used LIME to locate the most offensive five terms in an input comment. 1\% of the 200 comments we evaluated contained more than five offensive terms\footnote{Our limit of five was based on pragmatism. However, we acknowledge that we are unable to rule out sentences that have more than five offensive terms, and future work can investigate a higher limit (e.g., 10 or 15), or even finding all offensive terms.}. 
        
        While analysing the results of the modified GPS pipeline to paraphrase offensive comments, overall, the model provided good results (an average J score of 0.8) indicating high quality paraphrased suggestions that can replace offensive comments. We believe that the high quality is due to the stringent measures to compute the J score. When the suggestions produced were unable to meet the J score threshold, the seq2seq model was used, and this ensured that there was a paraphrased output for every offensive comment. The provided suggestions would make a writer think twice before posting an offensive comment. Thus, readers of these comments are less likely to be offended and this approach provides a pathway for the community to be free from toxic contents.

        \textbf{User evaluation of OLRS-SE}: Finally, in order to evaluate the utility of the proposed system, we conducted user evaluations using the implemented OLRS-SE, which was deployed as a web tool. The study collected feedback from 30 participants. The general responses of the participants were very positive about both the usability and the usefulness of the tool (see Section~\ref{subsection:user_evaluation_results}). However, the tool needs some improvements, such as the need to produce faster results, since only 46\% of the participants strongly agreed that the tool is faster compared to human judgments to detect offensive language (Q24). This is because the tool is somewhat slow to produce results due to the large number of ML models involved in the OLRS-SE pipeline. This is because some models are large neural network models (size up to 500 MB), which are slow to load into memory. Moreover, the tool runs on a shared cloud platform without GPUs. In addition, there were qualitative suggestions for speedy paraphrasing of offensive comments (the paraphraser takes an average of 4.6 seconds to rephrase a comment). These can be considered in future work by deploying our models in faster servers. The relevant files for the developed system are available in the link here\footnote{\url{https://github.com/jithincheriyan/Offense-detection-in-SE-communities}}.

        We would like to emphasize that the principal contribution of this work is the design, integration, and evaluation of the OLRS-SE system rather than the use of any particular language model. The proposed architecture is model-agnostic, with the classification component designed as a modular plug-and-play element that can readily accommodate advances in NLP. We used BERT as the primary model for many classification tasks as shown in Figure~\ref{fig:Cross validated results}, whose F1-Score ranges from 0.48 to 0.92. BERT was selected because it was a well-established and effective encoder at the time of research, providing a reliable baseline to evaluate the overall system. However, we acknowledge that the results of the OLRS-SE system have not been compared with the latest large language models such as GPT or Claude families of models. As newer encoder models (e.g., RoBERTa, DeBERTa, and ModernBERT) and large language models continue to evolve, these models can be integrated into the same architecture with minimal modification. Evaluating such replacements represents a natural extension of this work and would allow future studies to quantify the performance gains attributable to advances in language models while preserving the core contributions of the OLRS-SE framework.

        \textbf{Future work}: For some of the labels, the performance of the best model is less than desirable. For example, the F1-score for the target-detection model (Level-2 UN) was only 0.56. A key future work is to improve the performance of these models by collecting more training data (for deep learning models) and identifying effective sets of characteristics (for traditional models) to precisely detect such offences. In addition, we identified that the length of the comments varies between the various SE platforms included in this study. For example, GitHub comments were found to be lengthier than the other three, and Slack comments were observed to be shorter than the rest. This also might have contributed towards reduced ML result metrics. Important directions for future work include a systematic sensitivity analysis of setting thresholds for various metrics (e.g., LIME and J score) and quality evaluation of output suggestions from the seq2seq model.
        
        Another future work is to develop the tool as an extension to web browsers (e.g., as plug-ins) so that offensive language  can be detected across the internet spaces where rephrased suggestions can be provided, in order to make those places welcoming. Prompt-based offensive language reduction~\citep{kok_beyond_2022} is another potential strand to be considered as an extension to the system so that when an offensive comment is entered, the system would be able to provide some prompting messages such as \textit{Are you sure?}, along with the opportunity to rewrite the comment. A similar approach has been implemented in Tinder~\citep{noauthor_tinder_2021}. 
        
        Our work here can be used to investigate how offensive language use in software engineering projects can impact various activities in the life-cycle of software development. For example, the relationship between offensive language use during development and its impact on software developer productivity and quality (e.g., bugs) can be investigated. Also, our work enables scrutinizing the role of offensive language on hampering diversity and inclusion in software engineering communities (e.g., which diversity aspects are targeted by offenders such as gender and sexual orientation). OLRS-SE can be further extended to provide targeted messages explaining the reason for the exclusionary nature of the posts (e.g., message targeting gender). These are promising downstream research directions for our research.
        
        Our immediate strand of work is to make use of Large Language Models such as ChatGPT API to detect all the classes of offensive language taxonomy proposed in this work. In the work, we considered only BERT for offence classifications because that was the state of the art at the time the primary author worked on his PhD, and the rest of the transformer-based language models have emerged since then. 
        
        We contextualize and highlight our contributions in the related work section (Section \ref{sec:related_work}).

    \subsection{Threats to validity}
    \label{section:threats to validity}
         This section presents the threats to validity that can potentially affect the results reported in this work. 
     
    \textbf{Internal validity} - The foremost threat to internal validity is the trustworthiness of the various sub-systems that we used, for example, Regex and PAPI. We collected the complete set of Regex from SO, but not sure whether there can be anymore, or would catch all obfuscated offensive terms in English language. Similarly, the vulnerability of PAPI score is also a threat, that~\citep{hosseini2017deceiving} have questioned the integrity of PAPI score using adversarial examples. 
    
    Another threat is the linguistic barrier: we considered the comments generated in English only. However, users can post offensive posts in any language. Our proposed system is not able to detect such offence. Moreover, we have considered comments in textual form only, not in the form of emojis or other visual ways (e.g., GIFs) to represent or express offence. Also, while considering comments in English alone, the research is not able to detect offence with 100\% accuracy (as shown in our results) due to the nuanced nature of expressing language. Even though the ML models are trained on a reasonable amount of data, human creativity to use context-specific offensive language can fool systems put in place to reduce offence. In particular, this can be an important issue in detecting mild offence (i.e., UN offence) which can be subtle and nuanced.
        
    Another internal threat is the upper limit of 250 characters in the length of comments considered by this research. If someone expresses longer offensive contents, the system may fail to detect and classify the offence. This is due to the way our ML models are trained (which target short comments), even though the approach can be scaled in future work by considering longer text. However, we observed that only around 6\% of the total RO and UN comments were beyond that limit.  Another threat is the complexity of the  OLRS-SE pipeline. 
        
    Another threat is that we did not consider the context around the comment (e.g., whether the comment provided for a question or answer, the type of question and what other users have commented). Considering the context of commenting could be useful in accurately determining whether a comment is offensive. 
        
    \textbf{External validity} - The developed OLRS-SE has been found to provide promising results for comments from all the four platforms. However, the scalability of the system beyond these platforms needs to be evaluated before deploying the system as a decision support system for users of other SE platforms. Moreover, since the ML models for offence detection and classification are trained on SE-related datasets, the OLRS-SE pipeline cannot be directly used on other domains such as general social media platforms. Fine-tuning using relevant datasets and testing the system with comments from targeted social media platforms will be required.
    
    \textbf{Construct validity} - During Regex checks, only the first Regex match is counted, and with the presence of just one Regex match, the comment is classified according to the taxonomy. While we are confident that the comment is offensive in this case with the presence of just one offensive word, we did not consider all the offensive content within the comment to classify it as offensive or not. Although we do not believe this is a limitation, we need to test this approach more thoroughly. Having said that, for detecting all other levels (e.g., target identification), all words of the comment were considered.
     
        \begin{table}[h!]        
        \centering    
        \begin{tabular}{|c|c|c|c|c|c|c|c|c|c|c|c|c|}
            \hline
            
            \multirow{2}{1cm}{Work} & \multirow{2}{1.1cm}{Context/\\Domain} & \multicolumn{2}{c|}{Detection} & \multicolumn{7}{c|}{Classification} & \multirow{2}{1cm}{Expla\\nation} & \multirow{2}{1cm}{Rephr\\asing} \\  \cline{3-11}
            & & RO & UN & Indi. & Gene. & EA & SA & Othe. & Swea. & Racial& & \\ \hline

            \textcolor{blue}{Ours} & \textcolor{blue}{SE} & \textcolor{blue}\checkmark & \textcolor{blue}\checkmark &  \textcolor{blue}\checkmark &  \textcolor{blue}\checkmark &  \textcolor{blue}\checkmark & \textcolor{blue}\checkmark & \textcolor{blue}\checkmark & \textcolor{blue}\checkmark & \textcolor{blue}\checkmark& \textcolor{blue}\checkmark &  \textcolor{blue}\checkmark \\ \hline

            ~\citep{Raman_2020} & SE & \checkmark & X & X& X& X& X& X & X & X & X & X\\ \hline

           ~\citep{DBLP:conf/kbse/Sultana22a}& SE & X & X & X& X& X& X & X& \checkmark& X & X & X \\ \hline

           ~\citep{DBLP:conf/enase/Sayago-HerediaC22} & SE & \checkmark & X & X & X& X& X& X& X& X & X & X \\ \hline

           ~\citep{Ferreira_2022} & SE & \checkmark & \checkmark & X& X& X& X& X & X & X & X & X\\ \hline

           ~\citep{Sarker_2022} & SE & X  & X & X& X& X& X& X & X & X & X & X\\ \hline           

           ~\citep{Qiu_2022} & SE & \checkmark & X & X& X& X& X& X & X & X & X & X\\ \hline           

           ~\citep{Miller_did_you_2021} & SE & \checkmark & \checkmark & X& X& X& X& X & X & X & X & X\\ \hline

           ~\citep{Sarker_2023} & SE & \checkmark & X & X& X& X& X& X & X & X & X & X\\ \hline

        ~\citep{DBLP:journals/pacmhci/FerreiraCA21} & SE & \checkmark & \checkmark & X& X& X& X& X & X & X & X & X\\ \hline

        ~\citep{sarker2025landscape} & SE & X &X  & X& X& X& X& X & X & X & X & X\\ \hline

          ~\citep{ Toxispan_Sarker_2023} & SE & \checkmark &X  & X& X& X& X& X & X & X & \checkmark & X\\ \hline

          ~\citep{Explore_ChatGPT_Mishra_2024} & SE & \checkmark &X  & X& X& X& X& X & X & X &X & X\\ \hline   

           ~\citep{Fanton2021HumanintheLoopFD} & Non-SE & \checkmark & X & X& X& X& X& X& X & X & X & \checkmark \\ \hline

           ~\citep{zhu-bhat-2021-generate} & Non-SE & \checkmark & X & X& X& X& X& X & X & X & X & \checkmark  \\ \hline

           ~\citep{DBLP:journals/corr/abs-2005-12423} & Non-SE & X & X & X& X& X & X& X& X& \checkmark & X & \checkmark \\ \hline     

           ~\citep{zampieri-etal-2019-predicting} & Non-SE & \checkmark & X & \checkmark & \checkmark & X & X& \checkmark & X& X & X & X \\ \hline

            ~\citep{waseem2017understanding} & Non-SE & \checkmark & \checkmark & \checkmark & \checkmark & X& X& X & X & X & X & X\\ \hline

            \end{tabular}            
            \caption{Comparison of functionalities of offence detection and reduction across different research work}
        \label{Table:objectives_summary}
        \end{table}
        
        To identify the offensive comments from the SO dataset, we had set thresholds. For example, for detecting rude comments (RO), the PAPI score threshold was set to 0.7. This was based on the observation that the comments with PAPI scores beyond that threshold were deleted from SO. While setting such thresholds may be seen as arbitrary, we chose the limit so as to achieve binary classes (offensive or not). However, one may argue that offensiveness can be more nuanced (e.g., can lie in a continuum).  This can be remedied by developing a continuum of offence where the values for offence ranges from 0 to 1, which can then be compartmentalized into categories such as high, medium and low, similar to what has been proposed in other domains such as categorizing data waste~\citep{DBLP:conf/amcis/CorbettSL20}. Our work here only considers classifiers that cater for  high (RO) and low (UN) categories. More nuanced categorizations can be considered in the future work.

\section{Related work}
\label{sec:related_work} 
     
     This section compares the contributions of our current work with prior work in the SE domain on offensive language, focusing on the four phases of OLRS-SE framework (offence detection, classification, explanation and rephrasing). Detection refers to the presence or absence of offensive language within a text snippet. Offence classification is the process of detecting fine-grained attributes (such as the targets of abuse) in an offensive text~\citep{waseem-hovy-2016-hateful, inproceedings}. A fine-grained topology of abuse was proposed by Waseem et al.~\citep{waseem2017understanding}, that includes the type of abuse (explicit or implicit), possible targets of abuse (e.g., individual or group) and fine-grained classes of abuse (e.g., racism or sexism). Zampieri et al.~\citep{zampieri-etal-2019-predicting, zampieri-etal-2019-semeval} also extended the offence detection to target detection (individual or group targeted) in social media posts. These works include attributes in the above-mentioned fine-grained typology for the classification of offensive language.   We make three remarks pertaining to prior work in this arena.

     First, while there has been a large body of work focusing on offensive language detection on the internet~\citep{schmidt_survey_2017}, limited work investigated the nature and prevalence of such language in SE communities~\citep{DBLP:journals/ese/JongelingSDS17}. In this work, we bridge this gap by investigating the prevalence and nature of offensive language in online SE communities. 
     The presence and patterns of toxic interactions in GitHub issues were investigated by Raman et al.~\citep{Raman_2020}, and the impacts of such discussions are evaluated against code quality to correlate these~\citep{DBLP:conf/enase/Sayago-HerediaC22}. An assessment of Gitter comments using existing toxicity detectors also suggests that offence is present in these communities, however, none of the existing tools perform adequately to detect offence in SE specific discussions~\citep{Sarker_CORR}. However, ML models using language models such as BERT have been proposed as potential candidate models~\citep{Ferreira_2022, inproceedings_2}. This work thus adds to the growing body of literature showing that offensive language is present in all the four SE communities considered in this work. 
     
     Second, there has been some prior work on automated offence classification in SE communities. A taxonomy was proposed in~\citep{DBLP:conf/coin-ws/CheriyanSC20} to classify offence in SO comments, and a rule-based system was developed to detect swearing and racial comments in SO discussions~\citep{10.1145/3463274.3463805}. Following that, Sultana et al.~\citep{DBLP:conf/kbse/Sultana22a} developed ML models to detect sexist and misogynistic discussions in GitHub pull requests. Our work extends the classification scheme of~\citep{DBLP:conf/coin-ws/CheriyanSC20} by considering additional attributes in the presented taxonomy (type of abuse and target of abuse) and develops approaches to classify offensive language with the ultimate goal of reducing them. A recent work~\citep{DBLP:conf/ease/SavarimuthuZCYG23} has proposed a taxonomy of 11 attributes present in offensive language by manually analysing comments, and this is a fertile avenue for pursuing automatic classification of attributes in the future. 
     
    Thirdly, this is the first work that has attempted offence explanation. We explain to the poster why a comment is offensive and also propose non-offensive alternative suggestions, which is a novel proposition in the domain. Offence rephrasing has been considered by prior works~\citep{zhu-bhat-2021-generate, Krishna2020ReformulatingUS, chung-etal-2019-conan, DBLP:journals/corr/abs-2005-12423}, even though none of them are in SE context. The widely adopted strand of paraphrasing was generating counter narratives~\citep{zhu-bhat-2021-generate,chung-etal-2019-conan}, which has the limitation of needing an annotated dataset. So, we used language models to generate paraphrases for offensive text, which do not require annotated dataset. 

    Table~\ref{Table:objectives_summary} compares and summarises the contributions of this work against similar works in SE and non-SE domains based on the 11 specific functionalities achieved through OLRS-SE  (arranged in columns, where a tick mark represents the inclusion of a specific functionality). The last row shows that our work substantially extends prior work through developing approaches to detect, classify, explain and rephrase offensive language.  Many of the comparator works focus on only two out of 11 functionalities considered in our current work, both in SE and non-SE domains, thus highlighting the main contributions of this work.

\section{Conclusion}
\label{sec:conclusion}
    Researchers have reported that SE communities have become unwelcoming, based on judging the amount of offensive content available. Therefore, the goal of this work was to investigate the nature and prevalence of offensive language that exist in online SE communities, and develop a prototype system that can help reduce the amount of offensive language. By taking SO as a case study, we detected the nature of offences and quantified their prevalence. To develop ML models that can identify and classify the offences, we collected the deleted comment dataset of SO and built ML models. We developed six sets of 12 ML models (72 in total) to perform six tasks relevant to offence detection and classification. Using the best models from these six sets, we implemented the Offensive Language Reduction System (i.e., OLRS-SE), which can be used to reduce offensive language at source in SE communities. OLRS-SE was able to detect offences and classify them into fine-grained offence classes based on the developed taxonomy. OLRS-SE was able to highlight offensive words to the users and also paraphrase offensive posts into non-offensive ones. 
    We believe the developed system has the potential to reduce offensive language in online SE communities.

\appendix
\section{Additional explanation of the 11 steps in the paraphraser module}
\label{Sec: Appendix Paraphraser Module}

    The paraphraser module is implemented as 11 steps (in Figure 4 in the paper) with four main components: 1) SE relation checker, 2) LIME-based explainer, 3) modified GPS pipeline and 4) the sequence to sequence (seq2seq) model creation. The following sub-sections detail each of these main four components.            
    
        \subsection{SE relatedness check}
        \label{chapter9_subsection_SE_relation}
            The purpose of an SE relatedness check (Step 1) is to assess whether performing paraphrasing is worthwhile. If the comment is SE-related, toning-down the offensiveness of the comment preserves the disagreement nature and SE-relevance of the comment by keeping the negative sentiment and preserving the SE-related words of the comment (which is achieved through steps 2 to 9). Such a suggestion is more likely to be chosen by a user as a recommended suggestion instead of the original offensive comment. However, certain offensive comments which are not related to Software Engineering do not need to be paraphrased using our approach since they are merely offensive comments alone (i.e., containing expletives), without any relevance to Software Engineering. For example, the comment \textit{F*** you and your stupid ego}, is not related to Software Engineering. Since the offence is not specifically related to the SE domain, to remind the posters about the offensive nature of their comment, we developed a Sequence-to-Sequence (seq2seq) model (Step 10) that accepts the comments and generates a set of paraphrases using counter-narrative generation approach. This is described in detail in sub-section~\ref{subsection:seq2seq model}. Thus, we use two paraphrasing approaches, one for SE-related comments (steps 2-9) and the other (step 10) for non SE-related comments.
            
            To check how many offensive comments contain SE terms, we randomly selected 100 comments each from SO and Gitter offensive language datasets, and manually evaluated the SE relevance of these comments. We found that only 23\% of comments are related to SE and the rest are not related to SE. 
            
            The first step in the automated paraphrasing of SE-related offensive comments is to identify whether a comment is related to the SE domain. To achieve this, following Rahmi et al.~\citep{10.1145/3457784.3457796}, we built a corpus of SE words by collecting the most commonly occurring 5000 terms of SEthesaurus~\citep{DBLP:journals/tse/ChenCZX21}. SEthesaurus maps the morphological forms (abbreviations, misspellings and synonyms) of SE-related terms to their original SE text.  SEthesaurus was compiled from SO, which is a domain of interest for this work. We conducted the SE relatedness check with varying numbers of SEthesaurus terms (e.g., 1000, 2000, 5000 and 10000) and the corpus which achieved the best accuracy (29\%) on the above-mentioned dataset of 200 comments was the 5000 terms one. The difference in accuracy (6\%) is owing to the presence of false positive terms such as \textit{question} and \textit{answer} (29\%\textminus23\%) which can be present in non-SE related domains also. The SE corpus that we used in this work is available online\footnote{SE corpus: \url{https://doi.org/10.6084/m9.figshare.19441592}} as a JSON file, listing the SE terms and their corresponding number of occurrences in SEthesaurus.  
            
            The next sub-section describes the LIME-based explainer module which is subsequently used by the paraphraser module to remedy SE-related offensiveness.
   
        \subsection{LIME-based explainer module}
        \label{subsubsection_lime_explainer}
            If a comment has been identified as SE-related, the paraphraser needs to identify the most prominent words that contributed towards offensiveness in that comment. For that, we used a LIME-based explanation module that returns a list of the most prominent words that contributed towards offensiveness (Step 2). LIME~\citep{10.1145/2939672.2939778} is a widely used model-agnostic and post-hoc explanation method that locally explains ML model decisions.  LIME works with perturbed instances of inputs (i.e., comments). We set the number of perturbed instances of inputs to 500. Also, we set the number of features as 5, meaning that LIME identifies the most promising five words that caused offence. We limited the number of words to 5 since it was rare (in our experience) to find an offensive statement in the SO dataset with more than five offensive terms.

           For ML explanations using LIME, we used the \textit{Lime} package\footnote{LIME: \url{https://github.com/marcotcr/lime}} of Python. For a particular comment, the package returns a list of words in the comment and their corresponding scores as the explanation for a particular label (e.g., offensive label). The scores are the proportional probability contribution of a term towards the comment being assigned to a particular label. For example, for a particular offensive comment \textit{you and your stupid java answer}, LIME returns  the following list with the following top-5 terms: (stupid: 0.24732, your: 0.01414, you: 0.00514, answer: 0.00218, java: 0.00109). Since these scores are not normalized, we scale them to a range of 0 to 1 so that we are able to identify the most prominent tokens towards offensiveness (Step 3). Moreover, by setting a threshold, we were able to remove potential polite words that LIME may propose as impolite. After the scaling process, the above-mentioned output of LIME  will be: (stupid: 1, your: 0.05299, you: 0.01644, answer: 0.00442, java: 0). We set the LIME threshold as 0.5~\cite{Wattanakriengkrai_2022}. Tokens (i.e., terms) with scores higher than the threshold are selected for masking and the rest of the terms (potential impolite terms) are kept intact without masking. 
           
           Based on the approach described above, we cannot assume that all the offensive tokens of a comment will be identified by the LIME explanation module. The identification of the offensive words may vary according to the model behaviour on perturbed instances of the comment~\citep{10.1145/2939672.2939778}. Therefore, in order to ensure that all the offensive terms are identified for masking, we additionally perform a Regex check on the input comment using the identified Regexes from SO (Step 3). If any of the impolite terms are missed from the LIME-based explanation approach, we identify and mask those terms with \textit{[mask]} using Regex check.

            After identifying the most prominent words that contribute towards offensiveness in a comment, which can be shown to the user as the explanation for offence (e.g., by highlighting those words), we allow the comment to proceed to the next component of the paraphraser module, which is a modified GPS pipeline, and it is detailed next. The paraphraser finds replacement words for the masked offensive words thus reducing offence. 
        
        \subsubsection{Modified GPS pipeline}
        \label{chapter9_subsection_modified_GPS}
            
            The original GPS approach~\citep{zhu-bhat-2021-generate} employed a \textit{Generate-Prune-Select} pipeline that generates a large pool of candidate counter hate speech sentences for an offensive text. We modified the GPS system in three ways.

            First, we are not producing counter hate speech sentences for offensive comments since counter hate speech is a response system (in the form template sentences) that helps to reduce or avoid offensive language by providing non-negative argumentative feedback. Moreover, counter hate speech sentences might not retain the original non-offensive contents that are present within the offensive comment. In our modified GPS system, we retain the non-offensive contents of the original offensive comment. In other words, we keep the generated sentences that are close to the original comment that do not contain offensive language. 

            To generate a set of candidate paraphrases, we used RoBERTa~\citep{DBLP:journals/corr/abs-1907-11692}, a Masked Language Model (MLM), which is able to generate alternate tokens for a special class of token called \textit{[mask]}. The MLMs are originally trained on unsupervised data to generate contextual suggestions to fill the masked tokens. We use a multi-token RoBERTa model, which is a robust version of the BERT model, to generate multiple token suggestions for the [mask] tokens in the masked comment (Step 4). For the MLM model to generate sets of token suggestions, we fine-tuned the pretrained RoBERTa model using UN comments in an unsupervised mode so that the negative tone of the comment could be preserved. The outcome of the RoBERTa model is a set of token suggestions for each masked token. While RoBERTa is able to generate a large number of alternatives for a masked word, we limit our results to the top five token suggestions.

            Second, we prune not only the ungrammatical candidate sentences from the generator pool, but also the set of candidate paraphrases which do not retain the style or context of the original comment. For the original GPS pipeline, there is only one evaluation metric, which is grammatical accuracy. However, since grammatical accuracy cannot be considered as the sole metric to evaluate the quality of the generated paraphrases, we used a metric called J score~\citep{Krishna2020ReformulatingUS} which incorporates three aspects: style, context and grammatical accuracy for the evaluation of the generated paraphrases. 
            
            Third, for the selection phase, the original GPS pipeline uses a linear mapping of sentences into a vector space and then uses cosine similarity to determine the cosine distance from the original comment. Since semantic similarity alone cannot construct a valuable metric for detoxification evaluation~\citep{Krishna2020ReformulatingUS}, we used the J score as the metric to evaluate the paraphrased comments. We set a threshold for the J score (0.8 out of 1) to select the final set of suggestions and present them as recommendations to the writer. 
        
            The modified version of the GPS pipeline primarily includes similar components to those of the GPS pipeline: a generator that generates a set of candidate paraphrased suggestions for an input comment (Steps 4 to 6), a pruning system that prunes some unwanted suggestions (Steps 7) and finally a selection system to choose the best suggestions as recommendations to the user (Step 8).

            If none of the suggestions generated at Step 8 are able to meet the required J score threshold (Step 9), the GPS pipeline does not produce paraphrased suggestions. In that case, we make use of a seq2seq model that produces counter-narratives for an offensive comment. The next sub-section describes this seq2seq model. 
            
\begin{figure}[h!]    
            \centering            \includegraphics[width=\linewidth]{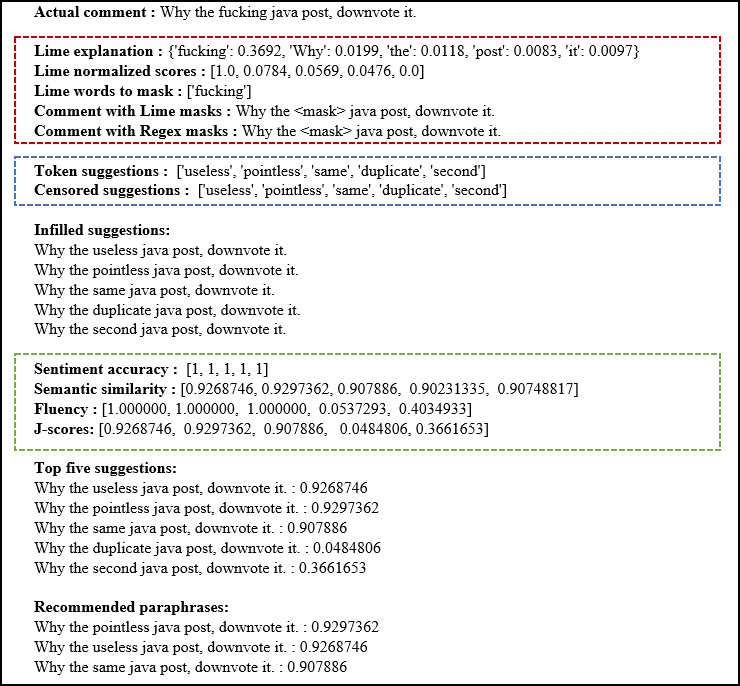}
            \caption{A screenshot of the modified GPS paraphrasing pipeline}
            \label{Fig:GPS_output_screenshot}
        \end{figure}

            A screenshot of the output of the paraphrasing pipeline using the modified GPS system is shown in Figure~\ref{Fig:GPS_output_screenshot}. For the highly offensive comment used as an example (i.e., \textit{Why the fucking java post, downvote it}), since the comment is SE-related, the LIME and Regex checks identify the word that contributes to the sentence being offensive (see results within the red box). The Generate module of the GPS pipeline generates alternative tokens for offensive tokens, and the tokens are censored using a Regex check (see results within the blue box). The results below the blue box are infilled suggestions where the masks are replaced with suggestions. The Pruning module calculates the J score of individual suggestions (see the results in the green box), as shown in the figure. Only the top three suggestions are recommended by the Select module and these paraphrased sentences are recommended according to the user (last set of lines in Figure~\ref{Fig:GPS_output_screenshot}).

        \subsection{Rephrasing non-SE-related comments}
        \label{subsection:seq2seq model}
            Similar to SE-related offensive comments, non-SE-related offensive comments should also be removed from the community discussions. In order to consider non-SE-related offensive comments for potential paraphrasing, we employ the counter-narrative approach in~\citep{chung-etal-2019-conan}. Thus, the objective of the seq2seq model is to generate counter-narratives for non-SE-related offensive comments (Step 10). 
            
            The main drawback of a manual counter hate speech generation approach is that a human is needed to actively engage users who may post hateful contents. Since it is not practically viable owing to the sheer volume of posts, a system that generates counter hate speech generation has gained attraction in online social settings~\citep{DBLP:journals/corr/abs-2005-12423, DBLP:conf/acl-alw/VidgenHGMBWBHT20}.  However, a system to automate counter hate speech generation must be trained with hand-curated data (called the parallel corpora)~\citep{chung-etal-2019-conan}, which is also a cumbersome task. This is because the parallel corpora must have source and corresponding target style sentences (i.e., offensive and corresponding non-offensive sentences). However, there are some initiatives in that direction. For example, CONAN\footnote{CONAN: \url{https://github.com/marcoguerini/CONAN}} is a large-scale, multilingual and expert-based dataset of Islamophobia comments, along with counter narratives to fight against that phenomenon.  
            
            Motivated by that approach, we developed a curated dataset of comments containing offensive language, along with counter-narratives for these comments. The dataset was prepared in such a way that for a source sentence (an offensive comment), there is a target sentence (a less offensive or non-offensive suggestion). The dataset contains 500 RO comments as the source sentences that are offensive and not related to SE discussions. The target set contains 500 sentences that are non-offensive comments. A seq2seq model trained on this dataset produces a counter sentence to an offensive sentence that does not have offensive language. This counter narrative when presented as an alternative to the offensive comment could be chosen by the authors and thus enable a more civilized conversation. For example, a comment such as \textit{f*** [that] question [was] allowed to stay open [but] mine get closed} can be rephrased as \textit{I am offended because my question got closed}. As mentioned above, we aim to preserve the negative tone of the offensive comment that expresses disappointment in the paraphrased suggestions, but with reduced offensiveness. The training data can be found online\footnote{Seq2seq model training data: \url{https://doi.org/10.6084/m9.figshare.19513705}}. 

            \begin{figure}[h!]    
            \centering            
            \includegraphics[width=\linewidth]{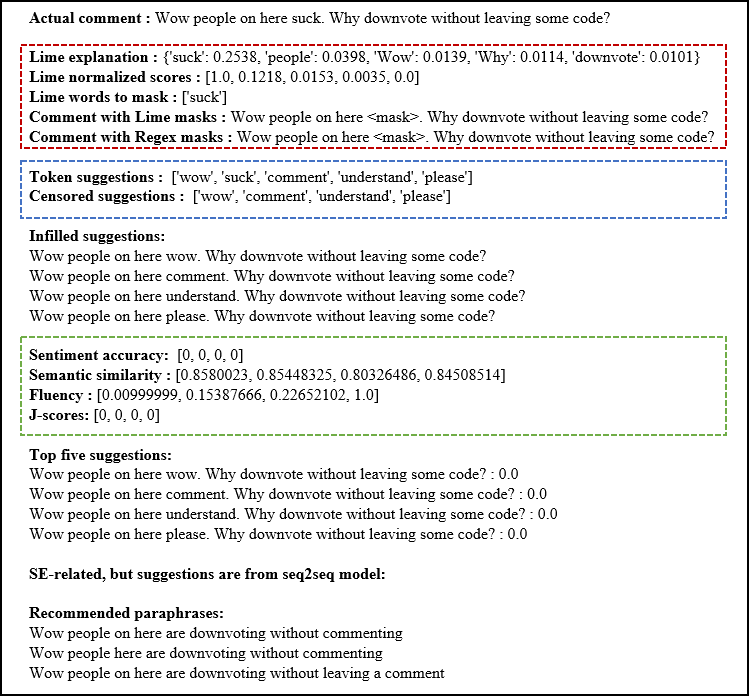}
            \caption{Screenshot of seq2seq paraphrasing of an SE-related comment}
            \label{Fig:se2seq Output}
        \end{figure}

            Figure~\ref{Fig:se2seq Output} shows a sample screenshot from the seq2seq model that paraphrases an offensive comment, which is SE-related. Here, the comment has been identified as SE-related and highly offensive. LIME and Regex modules identify the offensive words (see the red box), and the Generate module produces alternate tokens for the offensive comment (see the blue box). However, while calculating the J score, no suggestions were able to meet the required J score threshold (see the green box). Those suggestions (the second last set of lines) were removed as the scores are zeros (i.e., not considered as potential substitutions). Since there are no suggestions available, the  seq2seq model is used to generate a set of suggestions
            (last set of lines). This ensures that even though the modified GPS paraphrasing pipeline is unable to produce paraphrases for highly offensive comments, the seq2seq model will be able to achieve that. This was achieved  using the BART model, which was fine-tuned on the parallel corpora with offensive sentences as source statements and less offensive or non-offensive statements as target statements.


\bibliographystyle{elsarticle-num-names} 
\bibliography{JSW}

\end{document}